\documentclass{emulateapj}
\usepackage{amsmath}
\usepackage{graphicx}
\shorttitle{Velocity-space instabilities in a decreasing magnetic field}
\shortauthors{}

\begin{document}
\title{PIC Simulations of Velocity-Space Instabilities in a Decreasing Magnetic Field: Viscosity and Thermal Conduction}  

\author{Mario Riquelme\altaffilmark{1}, Eliot Quataert\altaffilmark{2} \& Daniel Verscharen\altaffilmark{3,} \altaffilmark{4}}
\altaffiltext{1}{Departamento de F\'isica, Facultad de Ciencias F\'isicas y Matem\'aticas, Universidad de Chile; mario.riquelme@dfi.uchile.cl}
\altaffiltext{2}{Astronomy Department and Theoretical Astrophysics Center, University of California, Berkeley, CA 94720; eliot@berkeley.edu}
\altaffiltext{3}{Mullard Space Science Laboratory, University College London, Dorking, Surrey, UK; d.verscharen@ucl.ac.uk}
\altaffiltext{4}{Space Science Center, University of New Hampshire, Durham, NH, USA}

\begin{abstract} 
\noindent We use particle-in-cell (PIC) simulations of a collisionless, electron-ion plasma with a decreasing background magnetic field, \textbf{\textit{B}}, to study the effect of velocity-space instabilities on the viscous heating and thermal conduction of the plasma. If $|\textbf{\textit{B}}|$ decreases, the adiabatic invariance of the magnetic moment gives rise to pressure anisotropies with $p_{||,j} > p_{\perp,j}$ ($p_{||,j}$ and $p_{\perp,j}$ represent the pressure of species $j$ [electron or ion] parallel and perpendicular to \textbf{\textit{B}}). Linear theory indicates that, for sufficiently large anisotropies, different velocity-space instabilities can be triggered. These instabilities in principle have the ability to pitch-angle scatter the particles, limiting the growth of the anisotropies. Our simulations focus on the nonlinear, saturated regime of the instabilities. This is done through the permanent decrease of $|\textbf{\textit{B}}|$ by an imposed plasma shear. We show that, in the regime $2 \lesssim \beta_j \lesssim 20$ ($\beta_j \equiv 8\pi p_j/|\textbf{\textit{B}}|^2$), the saturated ion and electron pressure anisotropies are controlled by the combined effect of the oblique ion firehose (OIF) and the fast magnetosonic/whistler (FM/W) instabilities. These instabilities grow preferentially on the ion Larmor radius scale, and make $\Delta p_e/p_{||,e} \approx \Delta p_i/p_{||,i}$ (where $\Delta p_j=p_{\perp,j} - p_{||,j}$). We also quantify the thermal conduction of the plasma by directly calculating the mean free path of electrons, $\lambda_e$, along the mean magnetic field, finding that $\lambda_e$ strongly depends on whether $|\textbf{\textit{B}}|$ decreases or increases. Our results can be applied in studies of low collisionality plasmas such as the solar wind, the intracluster medium, and some accretion disks around black holes.\end{abstract}


\section{Introduction}
\label{sec:intro}

\noindent  In low-collisionality plasmas, the change in the magnitude of the local magnetic field ($B\equiv |\textbf{\textit{B}}|$) generically drives a pressure anisotropy with $p_{\parallel,j} \ne p_{\perp,j}$ (where $p_{\perp,j}$ and $p_{\parallel,j}$ correspond to the pressure of species $j$ perpendicular and parallel to $\textbf{\textit{B}}$). This is a consequence of the adiabatic invariance of the magnetic moment of particles, $\mu_j \equiv v_{\perp,j}^2/B$, in the absence of collisions (where $v_{\perp,j}$ is the velocity of species $j$ perpendicular to $\textbf{\textit{B}}$).\newline

\noindent These pressure anisotropies can trigger various velocity-space instabilities, which are in principle expected to pitch-angle scatter the particles, to some extent mimicking the effect of collisions. The combined effect of pressure anisotropies and velocity-space instabilities can affect various large scale properties of the plasma, including its effective viscosity  \citep{SharmaEtAl06, SquireEtAl2017} and thermal conductivity \citep[see, e.g.,][]{RiquelmeEtAl2016, KomarovEtAl16}. This weakly collisional behavior is expected to be important in several astrophysical systems, including low-luminosity accretion flows around compact objects \citep{SharmaEtAl07}, the intracluster medium (ICM) \citep{SchekochihinEtAl05, Lyutikov07}, and the heliosphere \citep[][]{MarucaEtAl11,RemyaEtAl13}. \newline

\noindent In a previous work \citep{RiquelmeEtAl2016} we studied how the plasma viscosity and thermal conductivity are affected by an increase of $B$, which naturally drives $p_{\perp,j} > p_{||,j}$. In this paper we study the opposite case, where $B$ decreases and $p_{\perp,j} < p_{||,j}$. In this case several velocity-space instabilities can be excited, ultimately regulating the extent to which the $p_{\perp,j} < p_{||,j}$ anisotropy can grow. When only the electron dynamics is considered, two types of plasma waves are expected to be driven unstable by the electron pressure anisotropy: $i)$ the oblique electron firehose (OEF) modes, which are purely growing modes, and $ii)$ the Alfv\'en/ion-cyclotron (A/IC) modes, which are quasi-parallel, propagating waves, driven unstable by cyclotron-resonant electrons \citep{LiEtAl2000,CamporealeEtAl2008}.\footnote{Although the Alfv\'en/ion-cyclotron (A/IC) modes grow at wavelengths comparable to the electron Larmor radius, their name indicates that they correspond to the Alfv\'en branch that starts as low wavenumber ($k$) Alfv\'en mode and then becomes the ion-cyclotron mode at higher $k$.} Similarly, in presence of an ion pressure aniostropy $p_{\perp,i} < p_{||,i}$, there are also two types of modes that can grow unstable: $i)$ the oblique ion firehose (OIF) modes, which are purely growing modes, and $ii)$ the fast-magnetosonic/whistler (FM/W) modes, which are quasi-parallel, propagating waves, excited by cyclotron-resonant ions \citep{QuestEtAl1996,GaryEtAl1998,HellingerEtAl2000}.\footnote{Although gyrokinetic theory suggests that OIF and OEF modes correspond to the same Alfv\'en-mode branch \citep{Kunz2015,Verscharen2017}, we will consider them as separate instabilities, identifying them as two different growth-rate maxima in k-space.}\newline

\noindent In this work we studied the nonlinear, saturated properties of these instabilities making use of particle-in-cell (PIC) simulations. This is achieved by continuously decreasing the strength of the background magnetic field by externally imposing a shear motion in the plasma. This setup is interesting since in realistic astrophysical scenarios the pressure anisotropies are expected to be driven (via $B$ decrease) for a time significantly longer than the initial regime where the instabilities grow exponentially. \newline

\noindent Previous works have already studied this long term regime by simulating an expanding (instead of shearing) plasma. These works have used both hybrid-PIC simulations, which focused on the evolution of the ion anisotropy-driven instabilities \citep{MatteiniEtAl2006,HellingerEtAl2008}, and PIC simulations that mainly captured the role of electron anisotropy-driven modes \citep{CamporealeEtAl2010}. Thus, our work is intended to study the combined effect of the electron and ion pressure anisotropies on the nonlinear, saturated regime of the different unstable modes. This aspect of our study is motivated in part by previous linear dispersion analyses showing that the electron pressure anisotropy can significantly influence the evolution of both the FM/W and OIF modes \citep{MichnoEtAl2014,ManevaEtAl2016}.\newline

\noindent There are two important applications of our work. One is to quantify the so called ``anisotropic viscosity" of the plasma, which is controlled by the pressure anisotropies of the particles. This viscosity is believed to contribute significantly to the heating of electrons and ions in accretion disks and other low-collisionality plasmas \citep{SharmaEtAl06,SharmaEtAl07,SquireEtAl2017}. Also, the nonlinear evolution of the different velocity-space instabilities sets the pitch-angle scattering rate of electrons, which is key to determine their mean free path and, therefore, the thermal conductivity of the plasma. \newline

\noindent The paper is organized as follows. In \S \ref{sec:numsetup} we describe our simulation setup and strategy. In \S \ref{sec:interplay} we determine the saturated pressure anisotropy $\Delta p_j$ of ions and electrons, describing the responsible physical mechanisms. In \S \ref{sec:viscosity} we quantify the ion and electron heating. In \S \ref{sec:conduct} we measure the mean free path of electrons and ions, and determine their dependence on the physical parameters of the plasma.  In \S \ref{sec:conclu} we summarize our results and discuss their implications for various low-collisionality astrophysical plasmas.

\begin{deluxetable}{llllll} 
\tablecaption{Physical and numerical parameters of the simulations} 
\tablehead{ \colhead{Runs}&\colhead{$m_i/m_e$}&\colhead{$\omega_{c,e}/s$}&\colhead{$k_BT_e/m_ec^2$}&\colhead{N$_{\textrm{ppc}}$}&\colhead{L/R$_{L,e}$} } 
\startdata
  I1 &  $\infty$& 3600 &  0.28 & 40 & 210\\
  I2 &  $\infty$& 7200 &  0.28 & 40 & 210 \\
  I3 &  $\infty$& 7200 &  0.1 & 40 & 210 \\
  F1 &  64& 7200 &  0.28 & 40 & 640 \\
  F2 &  25& 7200 &  0.28 & 40 & 400 \\
  F3 &  25& 3600 &  0.28 & 40 & 400 \\
  F4 &  10& 7200 &  0.28 & 40 & 250
\enddata 
\tablecomments{A summary of the physical and numerical parameters of our simulations. These are the mass ratio $m_i/m_e$, the initial electron magnetization $\omega_{c,e}/s$, the ratio between electron temperature and rest mass energy, $kT_e/m_ec^2$, the number of particles per cell N$_{\textrm{ppc}}$ (including ions and electrons), and the box size in units of the typical initial electron Larmor radius $L/R_{L,e}$ ($R_{L,e} = v_{th,e}/\omega_{c,e}$, where $v_{th,e}^2=k_BT_e/m_e$, $k_B$ is Boltzmann's constant, and $T_e$ is the electron temperature). All of the runs initially have $\beta_i=\beta_e=2$ and $c=0.225 \Delta_x/\Delta_t$, where $\Delta_t$ is the simulation time step. The runs shown in this Table share the same electron skin depth $c/\omega_{p,e}/\Delta_x=5$ (where $\Delta_x$ is the grid point separation), but we used several other simulations to confirm numerical convergence by varying $c/\omega_{c,e}/\Delta_x$, N$_{ppc}$, and L/R$_{L,e}$.} 
\label{table:1D} 
\end{deluxetable} 

\section{Simulation Setup}
 \label{sec:numsetup}
\noindent We use the electromagnetic, relativistic PIC code TRISTAN-MP \citep{Buneman93, Spitkovsky05} in two dimensions. The simulation box consists of a square box in the $x$-$y$ plane, containing an initially isotropic plasma with a homogeneous initial magnetic field $\textbf{\textit{B}}_0$. We simulate a decreasing magnetic field by imposing a velocity shear given by $\textbf{\textit{v}} = -sx\hat{y}$, where $s$ is the shear rate of the plasma and $x$ is the distance along $\hat{x}$ ($\hat{x}$ and $\hat{y}$ are the unit vectors parallel to the $x$ and $y$ axes, respectively). From flux conservation, the $x$ and $y$ components of the mean field evolve as $d \langle B_x\rangle /dt =0$ and $d \langle B_y\rangle /dt = -s\langle B_x\rangle$. Thus, if $\langle B_x \rangle$ and $\langle B_y \rangle$ are positive, there will be a decrease of $\langle B_y \rangle$ and, therefore, of $|\langle \textbf{\textit{B}}\rangle |$. Therefore, we initially choose $\textbf{\textit{B}}_0\propto \hat{x} + 3.3\hat{y}$, which garantees a decrease of $|\langle \textbf{\textit{B}}\rangle |$ and a $p_{\perp,j} < p_{||,j}$ anisotropy during a simulation time $\sim 3s^{-1}$. \newline
 
\noindent By resolving the $x$-$y$ plane, our simulations can capture the quasi-parallel A/IC and FM/W modes, as well as the oblique OEF and OIF modes with their wave vectors $\textbf{\textit{k}}$ forming any angle with the mean magnetic field $\langle  \textbf{\textit{B}}\rangle$. The key parameters in our simulations are: the particle magnetization, quantified by the ratio between the initial cyclotron frequency of each species and  the shear rate of the plasma, $\omega_{c,j}/s$ ($j=i,e$), and the ion to electron mass ratio, $m_i/m_e$.  In typical astrophysical cases, $\omega_{c,j} \gg s$ and $m_i/m_e=1836$. Due to computational constraints, however, we will use values of $\omega_{c,j}/s$ and $m_i/m_e$ much larger than unity, but still much smaller than expected in real environments. This limitation will be taken into account when applying our simulation results to relevant astrophysical cases.\newline

\noindent Our simulations initially have $\beta_i = \beta_e = 2$ ($\beta_j \equiv 8\pi p_j/|\textbf{\textit{B}}|^2$). In almost all of our runs $k_BT_e/m_ec^2 = 0.28$, which implies $\omega_{c,e}/\omega_{p,e}=0.53$ (where $k_B$, $T_e$, and $\omega_{p,e}$ are Boltzmann's constant, the electron temperature, and the electron plasma frequency, respectively). We will change our simulation conditions by varying: $\omega_{c,e}/s$ and $m_i/m_e$ (which uniquely fix $\omega_{c,i}/s$ and $k_{B}T_i/m_ic^2$). Some of our simulations use ``infinite mass ions" (the ions are technically immobile, so they just provide a neutralizing charge), with the goal of focusing on the electron-scale physics.   These provide a useful contrast with our finite $m_i/m_e$ runs and allow us to isolate the impact of ion physics on the electrons. The numerical parameters in our simulations will be: N$_{\textrm{ppc}}$ (number of particles per cell), $c/\omega_{p,e}/\Delta_x$ (the electron skin depth in terms of grid size), $L/R_{L,i}$ (box size in terms of the initial ion Larmor radius for runs with finite $m_i/m_e$; $R_{L,i} = v_{th,i}/\omega_{c,i}$, where $v_{th,i}=\sqrt{k_BT_i/m_i}$), and $L/R_{L,e}$ (box size in terms of the initial electron Larmor radius for runs with infinite $m_i/m_e$). Table \ref{table:1D} shows a summary of our key runs. We ran a series of simulations ensuring that the numerical parameters (e.g., different N$_{\textrm{ppc}}$) do not significantly affect our results. Note that most runs used just for numerical convergence are not in Table \ref{table:1D}.\newline

\section{Pressure Anisotropies}
\label{sec:interplay}

\noindent In this section we focus on the nonlinear evolution of the ion and electron pressure anisotropies. As stated above, we will begin by showing simulations where ions have infinite mass. 

\subsection{Simulations with $m_i/m_e=\infty$}
\label{sec:whistlersonly}

\begin{figure}[t!]  \centering \includegraphics[width=9.8cm]{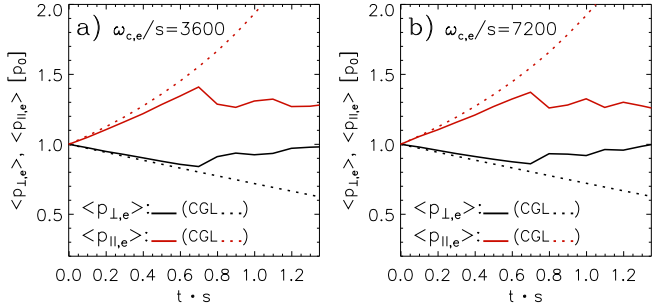}
  \caption{ The initial evolution of the electron pressures perpendicular ($p_{\perp,e}$; black-solid) and parallel ($p_{||,e}$; red-solid) to $\textbf{\textit{B}}$ for runs I1 and I2 in Table 1 (with $\omega_{c,e}/s = 3600$ and $\omega_{c,e}/s = 7200$, respectively). The black- and red-dotted lines show the expected $p_{\perp,e}$ and $p_{||,e}$ evolutions from the CGL or double adiabatic limit \citep{ChewEtAl56}. A significant deviation from adiabatic evolution can be seen at $t\cdot s \gtrsim 0.7$.}
\label{fig:cgl_3600and7200}
\end{figure}

\begin{figure*}[t!]  \centering \includegraphics[width=18cm]{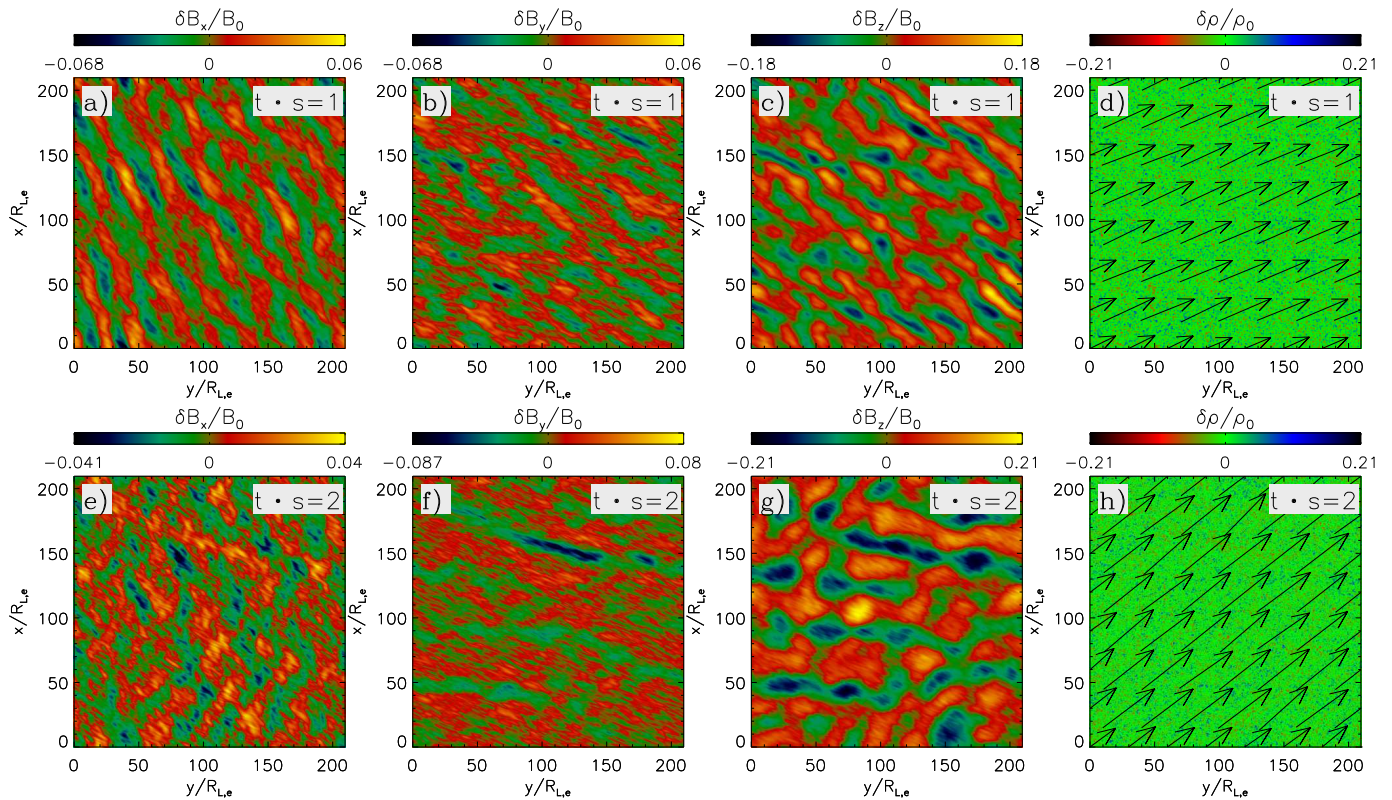}
  \caption{The three components of $\delta \textbf{\textit{B}}$ and plasma density fluctuations $\delta \rho$ at two different times: $t\cdot s = 1$ (upper row) and $t\cdot s = 2$ (lower row), for a simulation with infinite ion mass (run I1). Fields and density are normalized by the initial magnetic field, $B_0$, and the average density, $\rho_0$. The arrows in panels $d$ and $h$ show the magnetic field direction on the $x-y$ simulation plane. For this $m_i/m_e = \infty$ case, the magnetic fluctuations are dominated by the oblique OEF modes.}
\label{fig:fieldsbeta6}
\end{figure*}

\begin{figure}[t!]  \centering \includegraphics[width=8.7cm]{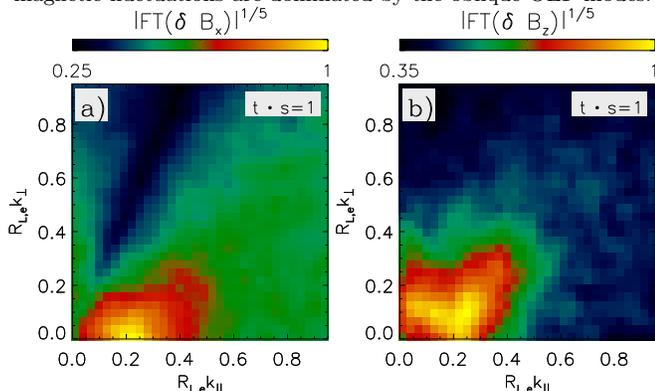}
  \caption{ The presence of the quasi-parallel A/IC and OEF modes for run I1 is shown by the magnitude of the Fourier transform of $\delta B_x$ (panel $a$) and $\delta B_z$ (panel $b$) at $t \cdot s = 1$. These quantities are plotted as a function of the wavenumbers parallel and perpendicular to $\langle \textbf{\textit{B}} \rangle$ ($k_{||}$ and $k_{\perp}$, respectively), are normalized by their maximum value, and are raised to the 1/5th power to provide better dynamical range. The contributions from (quasi-parallel) A/IC and (oblique) OEF modes with wavevectors satisfying $kR_{L,e} \sim 0.2$ are most clearly seen from panels $a$ and $b$, respectively.}
\label{fig:fftfirehose}
\end{figure}

\noindent Figure \ref{fig:cgl_3600and7200} shows the early time evolution (until $ t\cdot s \approx 1.3$) of the electron pressures perpendicular ($p_{\perp,e}$; black-solid line) and parallel ($p_{||,e}$; red-solid line) to $\textbf{\textit{B}}$ for runs I1 and I2 of Table \ref{table:1D}. These runs have infinite mass ions so that the electrons can only be affected by the electron anisotropy-driven OEF and A/IC instabilities, and their magnetizations are $\omega_{c,e}/s = 3600$ and $7200$, respectively. The black- and red-dotted lines show the expected evolutions of $p_{\perp,e}$ and $p_{||,e}$ from the CGL or double adiabatic limit \citep{ChewEtAl56}. We see that this adiabatic limit is reasonably well satisfied in the early stage of the simulations (until $t \cdot s \sim 0.7$), regardless of the magnetization $\omega_{c,e}/s$. After that, the growth of the electron anisotropy-driven instabilities provides enough pitch-angle scattering to stop the adiabatic evolution of the electron pressure.\newline

\noindent The presence of the OEF and A/IC instabilities can be seen from Figure \ref{fig:fieldsbeta6}, which shows the magnetic field fluctuations and plasma density in simulation I1. The upper row in Figure \ref{fig:fieldsbeta6} corresponds to $t\cdot s=1$, i.e., after one shear time, while the lower row corresponds to $t\cdot s=2$.   At all times the magnetic fluctuations are dominated by their $\delta B_z$ component, with wavenumbers $k$ ($\equiv |\textbf{\textit{k}}|$, where $\textbf{\textit{k}}$ is the wave vector) satisfying $kR_{L,e} \sim 0.2$, and with $\textbf{\textit{k}}$ being mainly oblique to the mean direction of $\textbf{\textit{B}}$. The presence of these oblique modes can also be seen from Figure \ref{fig:fftfirehose}$b$, which shows the Fourier transform of $\delta B_z$ at $t\cdot s=1$ as a function of the wavenumbers parallel and perpendicular to $\langle \textbf{\textit{B}} \rangle$ ($k_{||}$ and $k_{\perp}$, respectively). This suggests that the OEF modes contribute the most to the amplitude of the magnetic fluctuations. However, although smaller in amplitude, quasi-parallel A/IC modes can also be seen especially in the $\delta B_x$ component (see Figure \ref{fig:fieldsbeta6}$a$). This is also seen from Figure \ref{fig:fftfirehose}$a$, which shows that the Fourier transform of $\delta B_x$ (as a function of $k_{||}$ and $k_{\perp}$) is dominated by quasi-parallel modes.\newline

\begin{figure}[t!]  
\centering 
\includegraphics[width=9cm]{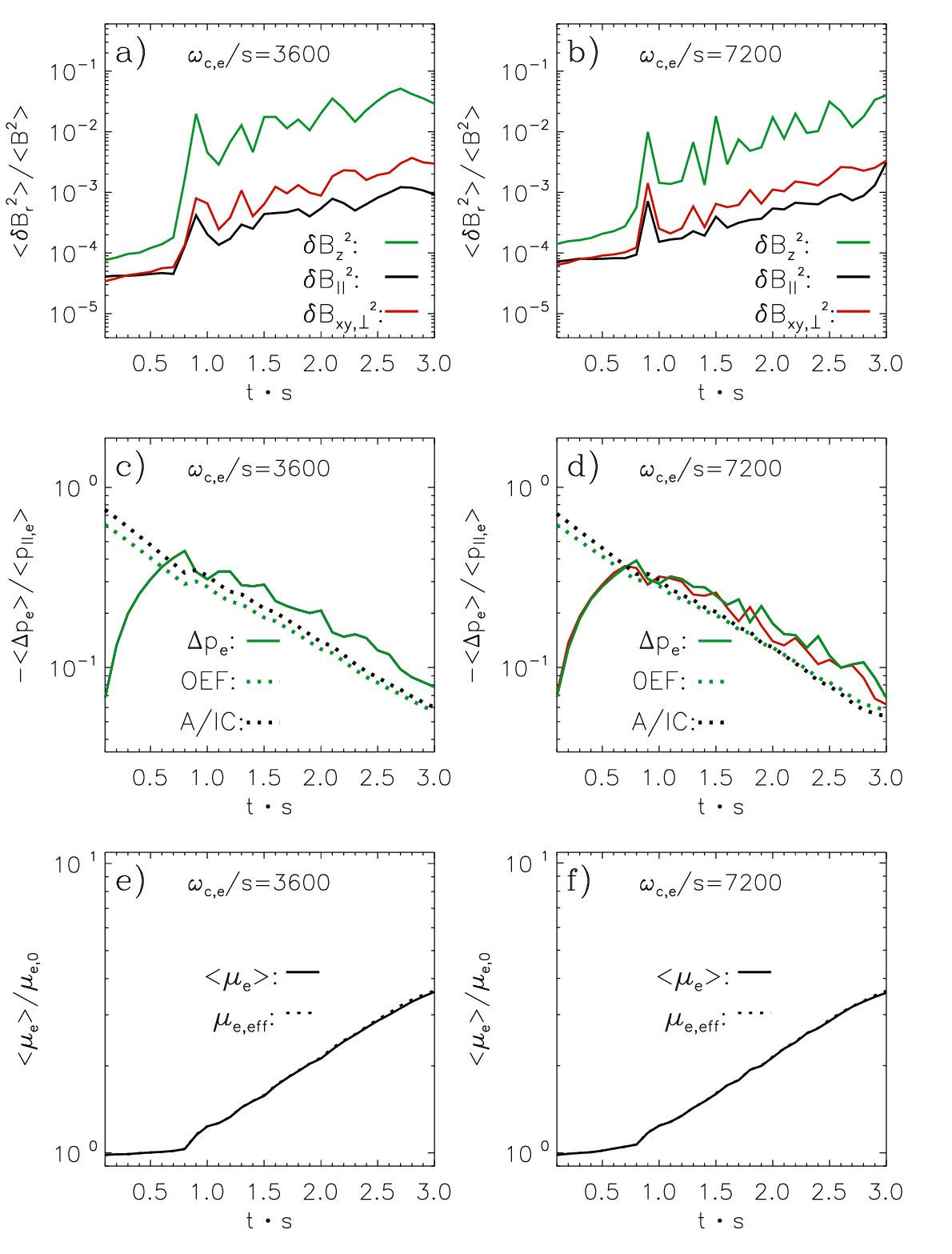}
\caption{The evolution of different volume-averaged quantities for two simulations with $\omega_{c,e}/s=3600$ (run I1; left column) and $\omega_{c,e}/s=7200$ (run I2; right column), which use $m_i/m_e=\infty$ (the ions simply provide a neutralizing charge). {\it Upper row:} the volume-averaged magnetic energy components, $\langle \delta B_r^2 \rangle$ pointing: along the $z$ axis ($\delta B_z^2$; green), parallel to $\langle \textbf{\textit{B}}\rangle$ ($\delta B_{||}^2$; black), and perpendicular to $\langle \textbf{\textit{B}}\rangle$ but in the $x$-$y$ plane ($\delta B_{xy,\perp}^2$; red), normalized by $\langle B^2\rangle$. {\it Middle row:} the evolution of the electron pressure anisotropy (green line), with the linear OEF and A/IC instabilities thresholds for growth rates $\gamma=s$ (dotted-green and dotted-black lines, respectively). The pressure anisotropy saturates at a value roughly consistent with the linear OEF and A/IC instabilities growing at the rate $\gamma = s$. Additionally, panel $d)$ shows in solid-red line the electron anisotropy for run I3, with the same parameters as run I2 but with $kT_e/m_ec^2=0.1$ instead of 0.28. The small difference between the solid-green and solid-red lines shows that the electrons being mildly relativistic should not affect substantially the evolution of $\Delta p_e$. {\it Lower row:} the electron magnetic moment; see equation \ref{eq:mu} and associated discussion for definitions of $\mu_e$ (solid) and $\mu_{eff,e}$ (dotted).} 
\label{fig:benandanis_wces1000and5000} 
\end{figure} 
\begin{figure*}[t!]
\centering
  \includegraphics[width=18cm]{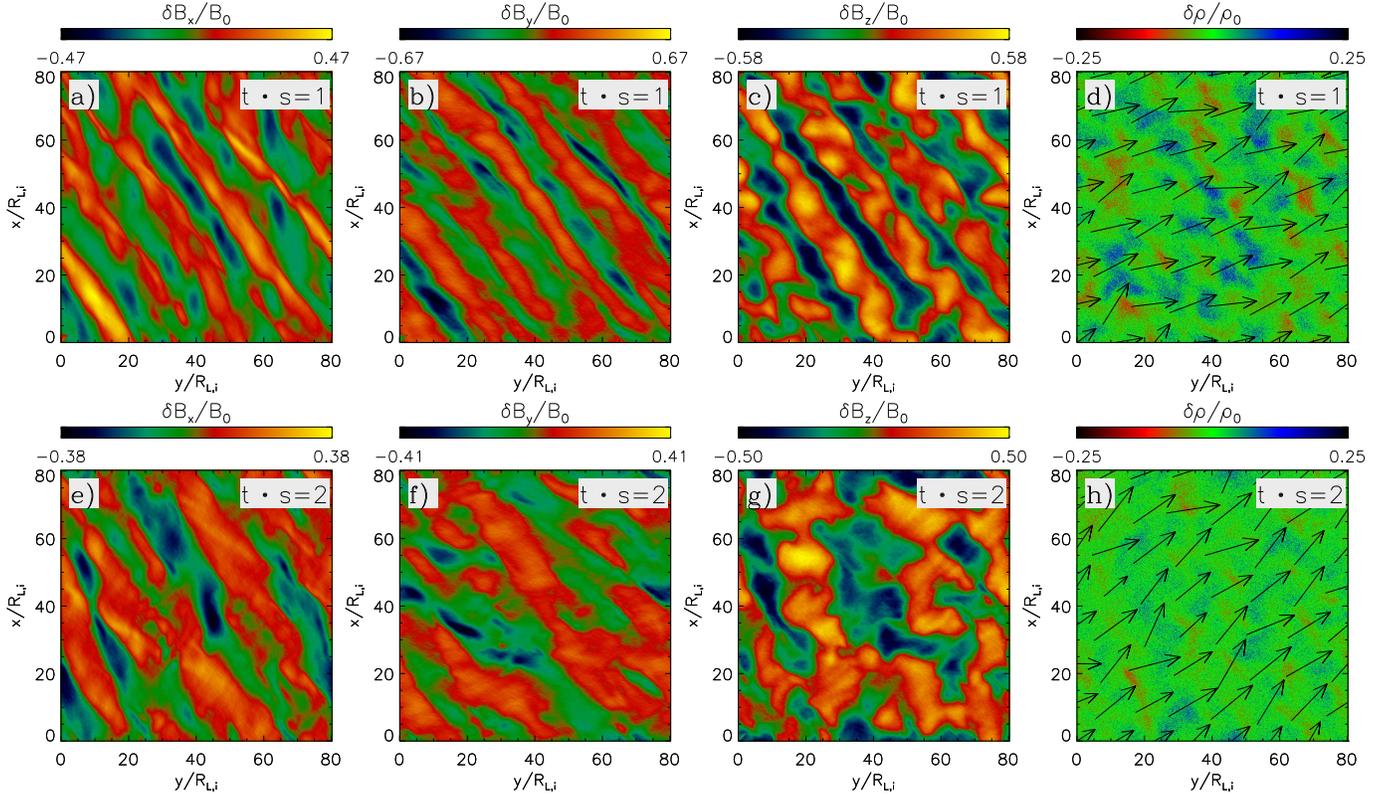}
  \caption{The three components of $\delta \textbf{\textit{B}}$ and plasma density fluctuations $\delta \rho$ at two different times: $t\cdot s = 1$ (upper row) and $t\cdot s = 2$ (lower row), for run F1 with $m_i/m_e = 64$. The fields and density are normalized by $B_0$ and the average density $\rho_0$, respectively. The arrows in panels $d$ and $h$ show the magnetic field direction on the $x-y$ plane. At both times, the magnetic fluctuations are dominated by a combination of the OIF and FW/M, with both modes contributing about the same energy. The OIF modes are oblique and appear mainly in the $\delta B_z$ component, the FM/W modes are quasi-parallel to $\langle \textbf{\textit{B}}\rangle$ and are apparent in the three axes.}
\label{fig:fldsmirrorwhistler}
\end{figure*}

\begin{figure}[t!]  \centering \includegraphics[width=8.7cm]{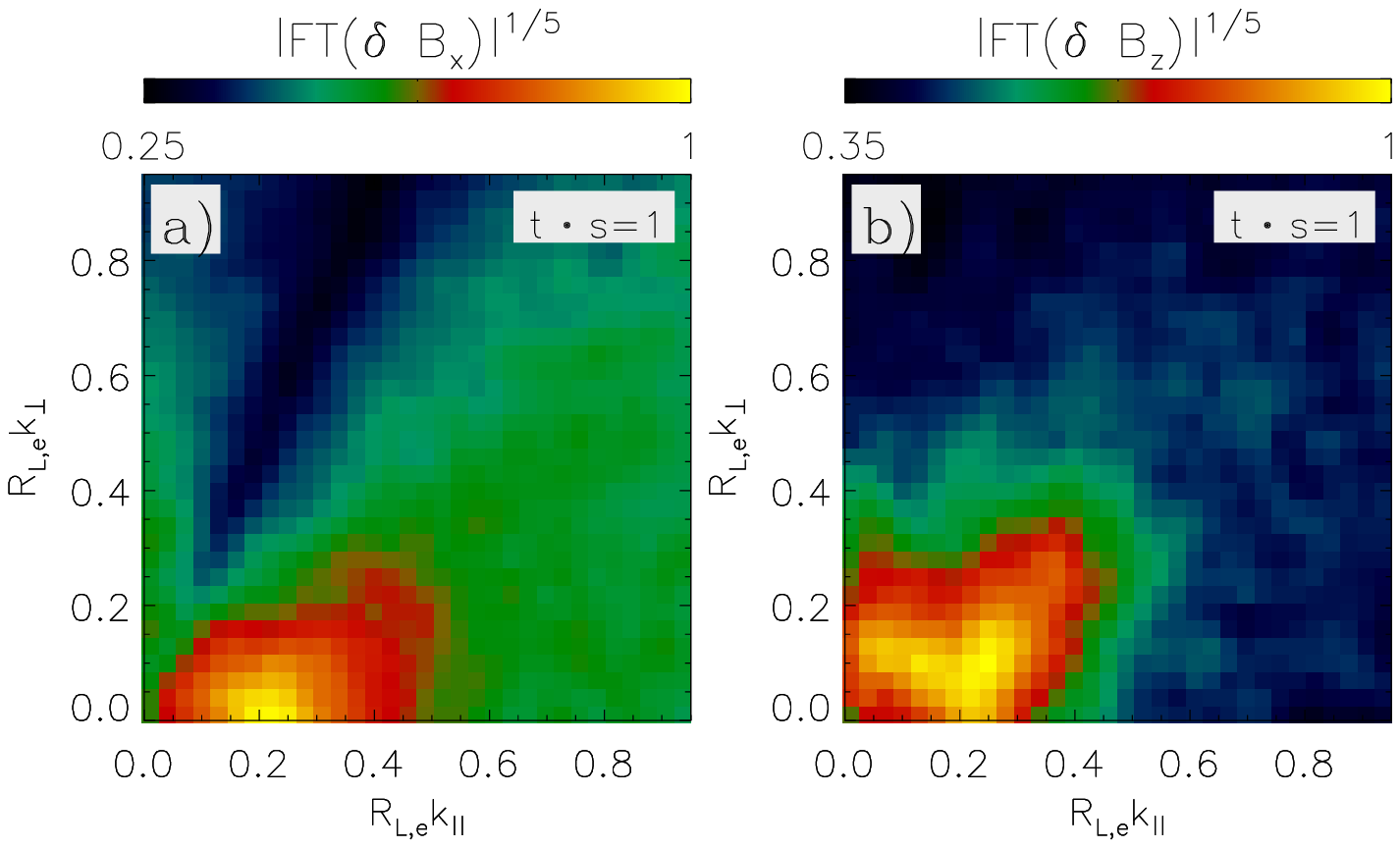}
  \caption{The magnitudes of the Fourier transform of $\delta B_x$ (panel $a$) and $\delta B_z$ (panel $b$) for run F1 ($m_i/m_e=64$) at $t \cdot s=1$, as a function of $k_{||}$ and $k_{\perp}$ and normailized by their maximum value (these quantities are raised to the 2/5th power to provide better dynamical range). The presence of the (quasi-parallel) FM/W modes with $kR_{L,e} \sim 0.4$ is clearly seen in both panels. The presence of the (oblique) OIF modes is apparent mainly in $\delta B_z$.}
\label{fig:fftfirehose_electronions}
\end{figure}

\noindent Figure \ref{fig:benandanis_wces1000and5000} shows the time evolution of the magnetic energy in $\langle \textbf{\textit{B}}\rangle$, the volume-averaged pressure anisotropy, and the electron magnetic moment for two runs, one with $\omega_{c,e}/s=3600$ and the other with $\omega_{c,e}/s=7200$ (runs I1 and I2 in Table \ref{table:1D}, respectively). Panels $c$ and $d$ show the volume-averaged pressure anisotropy $-\langle \Delta p_e\rangle /\langle p_{||,e}\rangle $ for these two runs, where $\Delta p_e=p_{\perp,e}-p_{||,e}$. For comparison, in both cases we plot the anisotropy threshold that would make the OEF and A/IC modes grow at a rate equal to the shearing rate, $s$. These thresholds were calculated using the linear Vlasov solver developed by \cite{Verscharen2013}. The calculations use mass ratio $m_i/m_e=1836$ and assume very cold ions ($\beta_i=10^{-4}$), which seeks to resemble our simulated $m_i/m_e=\infty$ situation where the ions only provide a neutralizing charge.\footnote{In order to make sure that using mildly relativistic electrons in our runs (where $k_BT_e/m_ec^2=0.28$) does not invalidate our comparison with the calculated thresholds (which assume non-relativistic electrons), in Figure \ref{fig:benandanis_wces1000and5000}$d$ we added in solid-red line the pressure anisotropy for run I3, which uses $k_BT_e/m_ec^2=0.1$ (while keeping the same $\omega_{c,e}/s$ and $\beta_e$). The very small difference between the two cases suggests that the effect of having mildly relativistic electrons is fairly small.} We see that the OEF and A/IC thresholds are quite similar, although with the OEF mode having a slightly lower threshold, especially in the case $\omega_{c,e}/s=3600$. This implies that both modes should play some role in regulating the electron anisotropy, with their relative importance depending weakly on the ratio $\omega_{c,e}/s$. Also, for both values of $\omega_{c,e}/s$ there is a reasonably good agreement between the electron anisotropy obtained from the simulations and the linear OFE and A/IC instability thresholds. This is thus consistent with the electron anisotropy being maintained at the level for the OEF and A/IC modes to grow at a rate close to $s$.\newline

\noindent The contribution of the different components of $\delta \textbf{\textit{B}}$ can be seen from Figures \ref{fig:benandanis_wces1000and5000}$a$ and \ref{fig:benandanis_wces1000and5000}$b$, which  show the magnetic energy along different axes as a function of time, normalized by the average magnetic energy in the simulation, $\langle  B^2\rangle /8\pi$. $\delta \textbf{\textit{B}}$ is decomposed in terms of $\delta B_z$ (component perpendicular to the simulation plane), $\delta B_{xy,\perp}$ (component parallel to the simulation plane but perpendicular to $\langle \textbf{\textit{B}} \rangle$), and $\delta B_{||}$ (component parallel to $\langle \textbf{\textit{B}} \rangle$). Clearly,  $\delta \textbf{\textit{B}}$ is dominated by its $z$ component (as already seen in Figure \ref{fig:fieldsbeta6}). This shows that, although the OEF and A/IC modes are expected to contribute to limiting the electron anisotropy, their contribution to the magnetic energy in $\delta \textbf{\textit{B}}$ is quite different. Indeed, our linear calculations show that, for the plasma parameters of runs I1 and I2, the OEF modes should satisfy $|\delta B_z|/(|\delta B_{||}| + |\delta B_{xy,\perp}|) \sim 4$. Thus the fact that in our runs $\delta B_z^2 \sim 10\delta B_{xy,\perp}^2$ implies that most of the magnetic energy is being contributed by the OEF modes. The quasi-parallel A/IC modes, which are most visible in the $\delta B_x$ component as can be seen from Figure \ref{fig:fieldsbeta6}$a$, make a significantly smaller contribution to $\delta \textbf{\textit{B}}$. Another sign of the dominance of the OEF modes is the growing and damping phases of $\delta B_z$ observed in Figures \ref{fig:benandanis_wces1000and5000}$a$ and \ref{fig:benandanis_wces1000and5000}$b$, which are likely related to the conversion of the saturated OEF modes into propagating waves that are rapidly damped through scattering with electrons, as it has been observed in previous initial value PIC simulations \citep[e.g.][]{HellingerEtAl2014}. The different contributions to $\delta \textbf{\textit{B}}$ are likely due to the slightly different anisotropy thresholds of the OEF and A/IC instabilities, as well as to the different amplitude at saturation expected for these two modes. Another possible factor is that the A/IC instability growth rate is very sensitive to the orientation of the pitch-angle gradients of the distribution function. Therefore, the A/IC instability can relax the distribution faster toward a stable configuration through pitch-angle scattering than the OEF instability.\newline

\noindent Finally, Figures \ref{fig:benandanis_wces1000and5000}$e$ and \ref{fig:benandanis_wces1000and5000}$f$ show the volume-average magnetic moment of the electrons, $\langle \mu_e \rangle$ ($\equiv \langle p_{\perp,e}/B \rangle$; black-solid line), for the same runs I1 and I2, respectively. It can be seen that until the onset of the OEF and A/IC exponential growth ($t\cdot s\approx 0.7$), $\langle \mu_e \rangle$ is fairly constant, implying the lack of efficient pitch-angle scattering. After that, $\langle \mu_e \rangle$ tends to increase at a rate close to the shear rate $s$. This implies the appearance of an effective pitch-angle scattering rate for the electrons, $\nu_{eff,e}$, due to their interaction with the OEF and A/IC modes. \newline

\noindent In order to help us to understand the way $\Delta p_j$ is regulated by the different velocity-space instabilities, we propose a second way to calculate the average magnetic moment of species $j$: 
\begin{equation}
\mu_{j,eff} \equiv \frac{\langle p_{\perp,j}\rangle }{\langle B\rangle }.
\label{eq:mu}
\end{equation}
This definition is useful because there can be cases where $\mu_{j,eff} \ne \langle \mu_j \rangle $. This is expected when, besides pitch-angle scattering, $\Delta p_j$ is partly regulated by relatively large fluctuations in $B$, which may spatially correlate with $p_{\perp,j}$ in a $\mu_j$ conserving way. This occurs, for instance, in the presence of large amplitude mirror modes \citep{Kunz2014,RiquelmeEtAl2015,RiquelmeEtAl2016}. Figures \ref{fig:benandanis_wces1000and5000}$e$ and \ref{fig:benandanis_wces1000and5000}$f$ show that, after $\langle \mu_{e} \rangle$ conservation is broken, $\mu_{e,eff}$ (dotted-black) and $\langle \mu_{e} \rangle$ (solid-black) are almost indistinguishable. This confirms that $\Delta p_e$ is regulated by an effective pitch-angle scattering provided by the OEF and A/IC modes, with no significant contribution from fluctuations in $B$.

\subsection{Simulations with finite $m_i/m_e$}
\label{sec:whistlerandmirror}

\noindent In order to study the effect of ions in regulating both the ion and electron pressure anisotropies, we now focus on simulations with finite ion to electron mass ratios $m_i/m_e$.  Since using $m_i/m_e \simeq 1836$ is computationally infeasible with our current resources, we have instead tried to ensure that our simulation results do not depend significantly on $m_i/m_e$, which is reasonably well achieved for $m_i/m_e = 25$ and 64. \newline

\noindent As an example, Figure \ref{fig:fldsmirrorwhistler} shows the three components of $\delta \textbf{\textit{B}}$ for run F1 of Table \ref{table:1D} ($m_i/m_e=64$ and $\omega_{c,e}/s=7200$).  The upper and lower rows correspond to $t\cdot s=1$ and $t\cdot s=2$, respectively. At both times, quasi-parallel and oblique modes are present with similar amplitudes, and with wavenumbers satisfying $kR_{L,i}\sim 0.4$ (where $R_{L,i}$ is the ion Larmor radius). While the quasi-parallel modes are apparent in the three components of $\delta \textbf{\textit{B}}$, the oblique modes appear mainly in $\delta B_z$. This can be seen more clearly from Figure \ref{fig:fftfirehose_electronions}, which shows the Fourier transform of $\delta B_x$ (Fig. \ref{fig:fftfirehose_electronions}$a$) and $\delta B_z$ (Fig. \ref{fig:fftfirehose_electronions}$b$) at $t\cdot s=1$ as a function of $k_{||}$ and $k_{\perp}$. The presence of quasi-parallel modes is clear in both panels, while the oblique modes mainly appear in $\delta B_z$. These features are consistent with the simultaneous presence of both OIF and FM/W modes. \newline

\noindent The presence or absence of fluctuations on electron Larmor radius scale is less clear from the Figures \ref{fig:fldsmirrorwhistler} and \ref{fig:fftfirehose_electronions}. We will come back to this question below.\newline 
\begin{figure*}[t!]
\centering
\includegraphics[width=19cm]{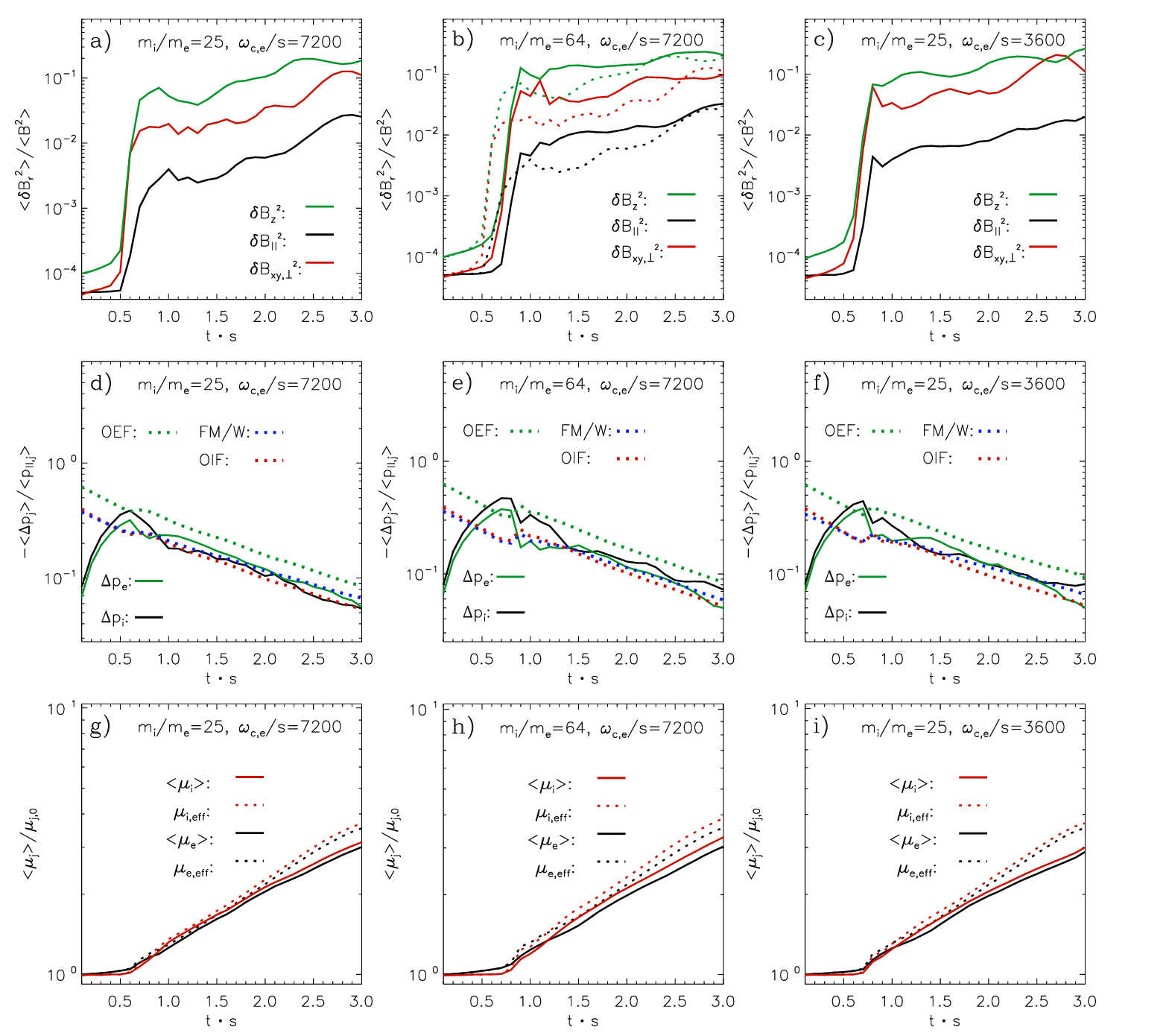}
\caption{Time evolution of volume-averaged quantities for simulations with $m_i/m_e=25$ and $\omega_{c,e}/s=7200$ (F2; left column), $m_i/m_e=64$ and $\omega_{c,e}/s=7200$ (F1; center column), and $m_i/m_e=25$ and $\omega_{c,e}/s=3600$ (F3; right column). The upper row shows the volume-averaged magnetic energy in three components of $\delta \textbf{\textit{B}}$: $\delta B_z$ (green), $\delta B_{||}$ (black), and $\delta B_{xy,\perp}$ (red), normalized by $B^2/8\pi$. The middle row shows the ion (black) and electron (green) pressure anisotropies, $\Delta p_j/p_{||,j}$, along with the anisotropy thresholds for the growth of different instabilities (using a growth rate of $s$). The dotted lines show thresholds assuming $\Delta p_e=\Delta p_i$, which correspond to the instabilities: OEF (dotted-green), OIF (dotted-red), and FM/W (dotted-blue). The solid-red and solid-blue lines also show the threshold for the growth of the OIF and FM/W instabilities, respectively, assuming $\Delta p_e=0$. Our results are consistent with the OIF and FM/W modes dominating the pressure anisotropies of ions and electrons, which satisfy $\Delta p_e \approx \Delta p_i$. The lower row shows the volume-averaged ion (solid-red) and electron (solid-black) magnetic moments, as well as the ``effective" averages defined as in equation \ref{eq:mu} for ions (dotted-red) and electrons (dotted-black), and normalized by the initial value of $\mu_j$.}
\label{fig:benandanis_wces2500_mime25and64}
\end{figure*}

\subsubsection{The Role of $m_i/m_e$}
\label{sec:rolemime}

\noindent Figure \ref{fig:benandanis_wces2500_mime25and64} compares the evolution of the energy in $\delta \textbf{\textit{B}}$, the ion and electron anisotropies, and $\mu_i$ and $\mu_e$ for simulations with different mass ratios and electron magnetization. The first and second columns compare simulations with the same electron conditions ($\omega_{c,e}/s=7200$, $k_BT_e=0.28 m_e c^2$, and $\beta_{e}=\beta_{i}=2$) but with different mass ratios: $m_i/m_e=25$ (run F2) and $m_i/m_e=64$ (run F1), respectively.\newline 
 
\noindent Figures \ref{fig:benandanis_wces2500_mime25and64}$d$ and \ref{fig:benandanis_wces2500_mime25and64}$e$ show the volume-averaged electron and ion pressure anisotropies as a function of time (green and black lines, respectively) for runs F2 and F1, respectively. These figures also show the anisotropy threshold for the growth of different instabilities, using a growth rate of $s$. In dotted lines we show thresholds that assume $\Delta p_e=\Delta p_i$, and correspond to the instabilities: OEF (dotted-green), OIF (dotted-red), and FM/W (dotted-blue). We do not include A/IC thresholds in this case. This is because our linear calculations show that the A/IC modes are subject to cyclotron-resonant ion damping when $T_i \sim T_e$, becoming stable in our runs with finite $m_i/m_e$ (this is true even for $m_i/m_e=1836$).\footnote{The A/IC modes are relevant for the simulations with fixed ions because these particles behave like very cold ions and can not damp the A/IC modes.} We see that our simulation results are fairly independent of the mass ratio, and can be summarized as follows: 
\begin{enumerate}
\item The ion and electron anisotropies evolve quite similarly in both cases. (This justifies using instability thresholds that assume $\Delta p_i=\Delta p_e$ for comparison.)
\item The obtained electron anisotropy is a factor $\sim 1.5$ smaller than the expected OEF threshold. 
\item The ion and electron anisotropies are best described by the thresholds of the OIF and FM/W instabilities with $\Delta p_i=\Delta p_e$.
\item The OIF and FM/W thresholds with $\Delta p_i=\Delta p_e$ are quite similar, which is consistent with the simultaneous presence of these modes in Figure \ref{fig:fldsmirrorwhistler}. 
\end{enumerate}
\noindent 

The fact that the electron anisotropy is close to the OIF and FM/W thresholds, and a factor $\sim 1.5$ smaller than the expected OEF threshold, shows that the OIF and FM/W modes are the ones with the largest effect on the electron anisotropy. This can be understood as due to the significant contribution of the electron pressure anisotropy to the growth of OIF and  FM/W modes. Indeed, our linear calculations show that the $\Delta p_i$ thresholds for the OIF and FM/W instabilities with $\Delta p_e=0$ are a factor of $\sim 2$ larger than in the $\Delta p_i=\Delta p_e$ case. This conclusion is also supported by the fact that, for the obtained electron anisotropy, the OEF modes are stable, indicating that the contribution from the OEF modes to the scattering of electrons is not expected to be important.\newline
 \begin{figure}[t!]
  \centering
  \includegraphics[width=10.2cm]{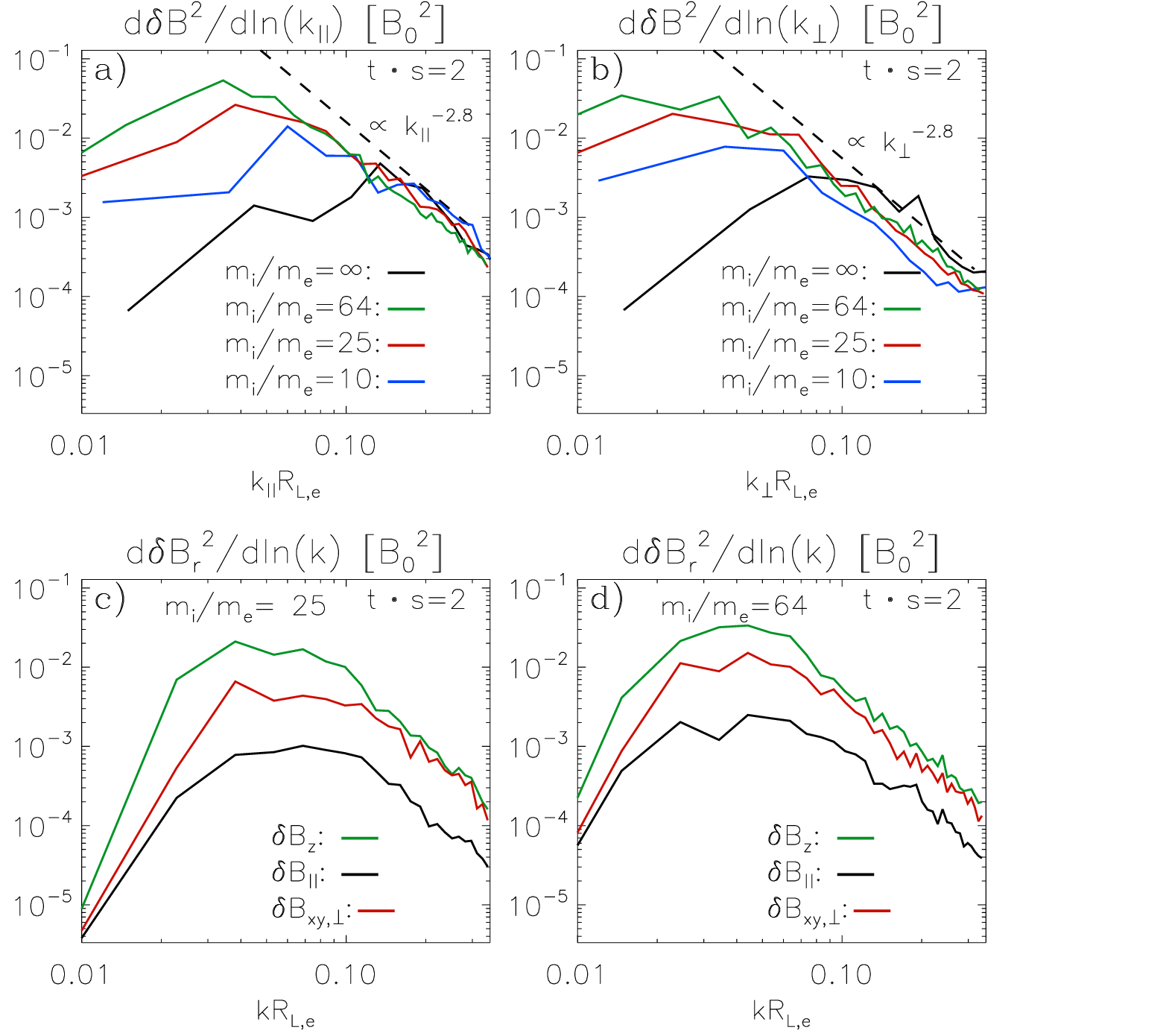}
  \caption{{\it Panel a:} the magnetic energy per logarithmic interval of $k_{||}$ is plotted at $t\cdot s=2$ for runs with $m_i/m_e=10, 25, 64,$ and $\infty$ (runs F4, F2, F1, and I2, respectively; $k_{||}$ is the magnitude of the wave vector component parallel to $\langle \textbf{\textit{B}} \rangle$. {\it Panel b:} the same as in panel $a$ but for $k_{\perp}$. {\it Panel c:} the power spectra for run F2 ($m_i/m_e=25$) by components: $\delta B_z^2$, $\delta B_{||}^2$, and $\delta B_{xy,\perp}^2$, and in terms of $k$ ($k^2=k_{||}^2+k_{\perp}^2$). {\it Panel d:} the same as in panel $c$ but for run F1 ($m_i/m_e=64$).}
\label{fig:spectra}
\end{figure}
 \begin{figure}[t!]
  \centering
  \includegraphics[width=9.2cm]{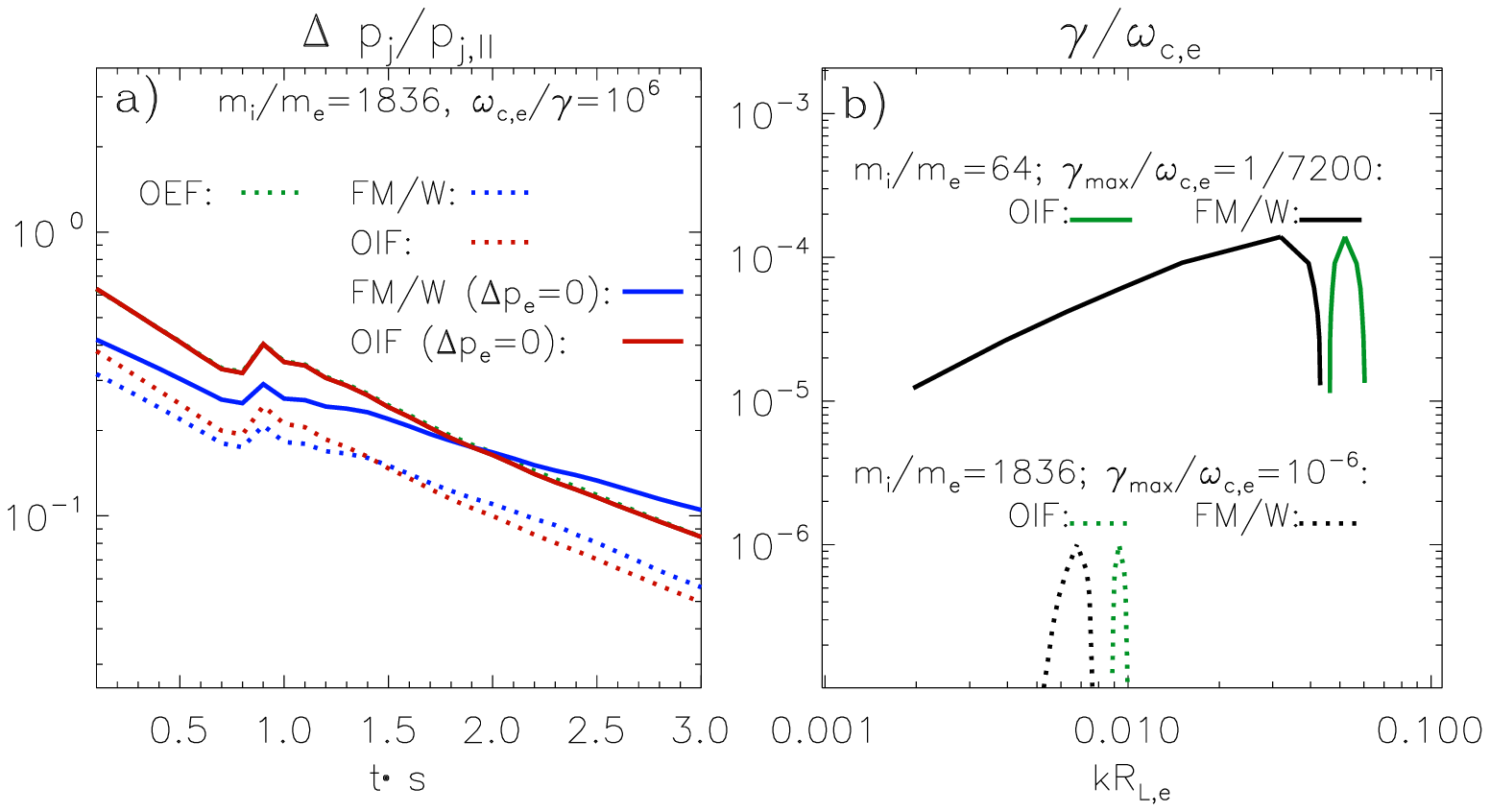}
  \caption{{\it Panel a:} Calculated pressure anisotropy thresholds for different instabilities in the case $m_i/m_e=1836$ and $\omega_{c,e}/\gamma=10^6$, using the Vlasov solver of \cite{Verscharen2013} and assuming the same $\beta_j$ evolution of our simulations ($\gamma$ represents the growth rate for the different instabilities). The cases $\Delta p_e=\Delta p_i$  are shown in dotted lines and $\Delta p_e=0$  in solid lines. The OEF, OIF, and FM/W instabilities are represented by green, red, and blue colors. For these more realistic parameters, the OIF and FM/W instabilities are expected to continue to dominate in the regulation of both ion and electron anisotropies, rendering $\Delta p_e\approx \Delta p_i$. {\it Panel b:} The growth rate $\gamma$ as a function of $kR_{L,e}$ for OIF and FM/W modes (green and black lines, respectively), assuming $\beta_i=\beta_e=10$, $\Delta p_e=\Delta p_i$. We consider two regimes: $i)$ $m_i/m_e=64$ with $\Delta p_j$ such that the maximum growth rate $\gamma_{max}= \omega_{c,e}/7200$ (solid lines), and $ii)$ $m_i/m_e=1836$ with $\Delta p_j$ such that $\gamma_{max}= \omega_{c,e}/10^{6}$ (dotted lines). We choose the maximum and minimum $|{\bf k}|=k$ for each $\gamma$.}
\label{fig:growthrates}
\end{figure}

\noindent Finally, we performed similar linear threshold calculations for the case $m_i/m_e=1836$ with $\omega_{c,e}/\gamma=10^6$ (where $\gamma$ represents the growth rate of the different instabilities), which are shown in Figure \ref{fig:growthrates}$a$. We find that the thresholds of the OIF (dotted-red) and FM/W (dotted-blue) modes with $\Delta p_e=\Delta p_i$ continue to be similar and smaller than the OEF (dotted-green) threshold by a factor $\sim 1.5$ (the dotted-green line almost coincides with the solid-red line). Also, the OIF and FM/W thresholds with $\Delta p_e=\Delta p_i$ are $\sim 1.5$ times smaller than in the $\Delta p_e=0$ case. This implies that, for realistic mass ratios and magnetizations, the dominant instability for the regulation of ion and electron anisotropies should continue to be the OIF and FM/W instabilities.\footnote{Although the OIF and FM/W thresholds are similar, there is the trend for the FM/W threshold to be smaller than the OIF threshold at early times ($t\cdot s\lesssim 1.8$, $\beta_i \lesssim 7$), while the opposite situation happens at late times ($t\cdot s\gtrsim 1.8$, $\beta_i \gtrsim 7$). This implies that in the more realistic cases there could be a clearer dominance of the FM/W (OIF) modes for $\beta_i \lesssim 7$ ($\gtrsim 7$). This is consistent with the OIF dominance shown in the hybrid-PIC simulations with fluid electrons presented by \cite{Kunz2014}, which use $\beta_i \sim 100$ and where the ions are significantly more magnetized than in our simulations.}\newline 

\noindent Figures \ref{fig:benandanis_wces2500_mime25and64}$a$ and \ref{fig:benandanis_wces2500_mime25and64}$b$, show the magnitude of the volume-averaged magnetic energy in the same three components of $\delta \textbf{\textit{B}}$ plotted in Figure \ref{fig:benandanis_wces1000and5000}: $\delta B_z$ (solid-green), $\delta B_{||}$ (solid-black), and $\delta B_{xy,\perp}$ (solid-red), for $m_i/m_e=25$ and 64, respectively (for comparison the case $m_i/m_e=25$ is replicated in Figure \ref{fig:benandanis_wces2500_mime25and64}$b$ using dotted lines). For the two mass ratios, the two components perpendicular to $\langle \delta \textbf{\textit{B}}\rangle$ dominate, with $\langle \delta B_z^2 \rangle$ being most of the time $\sim 3$ times larger than $\langle \delta B_{xy,\perp}^2 \rangle$. This result implies that the OIF and FM/W modes are contributing comparable energy to $\delta \textbf{\textit{B}}$, as was also noticeable from Figure \ref{fig:fldsmirrorwhistler}. Indeed, our linear calculations show that for the OIF modes $\delta B_z^2 \gg \delta B_{xy,\perp}^2$, while for the FM/W modes $\delta B_z^2 \sim \delta B_{xy,\perp}^2$,\footnote{Indeed, using the Vlasov solver of \cite{Verscharen2013} one can obtain that, for the parameters of runs F1 ($m_i/m_e=64$) and F2 ($m_i/m_e=25$), the OIF modes satisfy $|\delta B_z|/(\delta B_{xy,||}^2+\delta B_{xy,\perp}^2)^{1/2} \sim 5$.} which implies that most of the $\delta B_{xy,\perp}$ component is being produced by FM/W modes. \newline

\noindent The amplitudes of the OIF and FM/W modes appear to depend on the mass ratio. Although time dependent, the magnitude of $\delta B_z^2$ in the $m_i/m_e=64$ case is on average $\sim 1.5$ times larger than in the $m_i/m_e=25$ case. Since the magnetizations $\omega_{c,i}/s$ of the two runs differ by a factor $\sim2.6$ ($=64/25$), this is roughly consistent with previous studies of the OIF instability that show that $\delta B^2$ at saturation should scale as $\delta B^2/B^2\propto (s/\omega_{c,i})^{1/2}$ \citep{Kunz2014}. On the other hand, $\delta B_{xy,\perp}^2$ in the $m_i/m_e=64$ case is about $\sim 2$ times larger than for $m_i/m_e=25$, which is roughly consistent with the expectation for the FM/W modes to have a saturation amplitude that satisfies $\delta B^2/B^2\propto s/\omega_{c,i}$. This scaling can be obtained from the expected effective ion scattering frequency by resonant waves, $\nu_{eff,i}$, which scales as
\begin{equation}
\nu_{eff,i} \sim \frac{\delta B^2}{B^2}\frac{\omega_{c,i}^2}{k_{||}v_{||}},
\end{equation}
where $k_{||}$ and $v_{||}$ are the wave vector component and the particle velocity component parallel to $\langle \textbf{\textit{B}}\rangle$ \citep{Marsch2006}. For the case of the quasi-parallel FM/W waves, we obtained from the simulations that $k_{||}R_{L,i} \sim 0.3$, meaning that for most particles $k_{||}v_{||} \propto \omega_{c,i}$ and that $\nu_{eff,i} \propto (\delta B^2/B^2)\omega_{c,i}$. Thus, since at FM/W saturation one expects $\nu_{eff,i} \sim s\cdot p_{||,i}/\Delta p_i$ (see Equation \ref{eq:scaling} below), this implies that $\delta B^2/B^2\propto s/\omega_{c,i}$ (considering that the change in $\Delta p_i/p_{||,i}$ between the $m_i/m_e=25$ and 64 cases is small).\newline


\noindent Finally, in panels \ref{fig:benandanis_wces2500_mime25and64}$g$ and \ref{fig:benandanis_wces2500_mime25and64}$h$ we compare $\langle \mu_j\rangle  =\langle p_{\perp,j}/B\rangle $ and $\mu_{j,eff} =\langle p_{\perp,j}\rangle /\langle B\rangle $ for both ions and electrons.  We see that the change in $\mu_{j,eff}$ tends to be somewhat larger than the one in $\langle \mu_j\rangle $ (by $\sim 20\%$) for the two mass ratios tested. This implies that the combined effect of the OIF and FM/W modes reduces $p_{\perp,j}$ in a way that mainly breaks the adiabatic invariance of $\mu_j$, with the preservation of $\mu_j $ due to changes in the field configuration playing a small role.\newline

\subsubsection{The Role of $\omega_{c,e}/s$}
\label{sec:rolewces}

\noindent It is also important to understand the role of electron magnetization, $\omega_{c,e}/s$, while keeping the same mass ratio. This can be done by looking at the first and third columns in Figure \ref{fig:benandanis_wces2500_mime25and64}, which compares simulations with $m_i/m_e=25$ but with electron magnetization $\omega_{c,e}/s=7200$ and 3600 (runs F2 and F3 in Table \ref{table:1D}, respectively). We see that the two runs reproduce essentially the same results in terms of $\mu_j$ and $\Delta p_j$ evolution, with the only difference being in the amplitude of the magnetic fluctuations. Apart from some significant time variability, the amplitude of the $\delta B_z$ and $\delta B_{xy,\perp}$ components in the $m_i/m_e=25$, $\omega_{c,e}/s=3600$ run are quite similar to the case $m_i/m_e=64$, $\omega_{c,e}/s=7200$. Since these two runs have a very similar ratio $\omega_{c,i}/s$ ($=144$ and $113$, respectively), this result is consistent with the dependence of the OIF and FM/W saturated amplitude on $\omega_{c,i}/s$ mentioned in \S \ref{sec:rolemime}.\newline

\subsubsection{Breaking of $\mu_e$ Adiabatic Invariance}
\label{sec:pitchangle}

\noindent An important question is whether the ion-scale instabilities alone are capable of explaining the break in the electron magnetic moment shown in Figure \ref{fig:benandanis_wces2500_mime25and64}$g$, \ref{fig:benandanis_wces2500_mime25and64}$h$, and \ref{fig:benandanis_wces2500_mime25and64}$i$ (which starts at $t\cdot s \sim 0.7$). \newline

\noindent We explore this issue by comparing the power spectra of the fluctuations in our finite $m_i/m_e$ runs with the power spectrum produced in the case with $m_i/m_e=\infty$. This is done in Figures \ref{fig:spectra}$a$ and \ref{fig:spectra}$b$, where the magnetic energy per logarithmic unit of $k_{||}$ and $k_{\perp}$ is plotted at $t\cdot s=2$ for runs with $m_i/m_e=10, 25, 64,$ and $\infty$ (runs F4, F2, F1, and I2, respectively; $k_{||}$ and $k_{\perp}$ are the magnitude  of the wave vector components parallel and perpendicular to $\langle \textbf{\textit{B}} \rangle$, respectively). The electrons in these simulations have the same conditions ($k_BT_e/m_ec^2=0.28$, $\omega_{c,e}/s=7200$, and initial $\beta_e=2$), so the different results are only due to the different $m_i/m_e$. We see that:
\begin{enumerate}
\item In the cases with finite mass ratio, as $m_i/m_e$ increases the peaks of the spectra shift to longer wavelengths (in units of $R_{L,e}$) in a way consistent with the growth of the $R_{L,i}/R_{L,e}$ ratio.
\item In the same way, as $m_i/m_e$ increases there is a growth in the amplitude of the peak of the spectra, which accounts for the expected increase in the amplitude of the OIF and FM/W modes as $\omega_{c,i}/s$ decreases. 
\item The energy of the magnetic fluctuations on scales of $k_{\perp} R_{L,e} \sim k_{||}R_{L,e}\sim 0.2$ is quite similar regardless of the used mass ratio. 
\item For finite mass ratios, the power spectra develop, via power cascade, a tail that behaves as:\footnote{This behavior differs from the one obtained from hybrid-PIC simulations \citep{Kunz2014}, where $d\delta B^2/d\textrm{ln}(k_{||}) \propto k_{||}^{-3.8}$ and $d\delta B^2/d\textrm{ln}(k_{\perp}) \propto k_{\perp}^{-3.8}$. This possibly denotes the influence of the electrons in the cascading process.} $d\delta B^2/d\textrm{ln}(k_{||}) \propto k_{||}^{-2.8}$ and $d\delta B^2/d\textrm{ln}(k_{\perp}) \propto k_{\perp}^{-2.8}$.  
\end{enumerate}

\noindent The similar amplitude of the magnetic fluctuations on electron scales suggests that the break in the adiabatic invariance of $\mu_e$ can in principle be caused by the OIF and FM/W instabilities via a three steps scenario: $i)$ ion and electron anisotropies create magnetic fluctuations through the OIF and FM/W instabilities, $ii)$ part of the energy in the magnetic fluctuations is transferred to electron scales via power cascade, and $iii)$ electrons are pitch-angle scattered by these fluctuations producing the break in $\mu_e$ invariance. We can compare the contributions from the OIF and FM/W modes to the cascading process by looking at the power spectra by components: $\delta B_z^2$, $\delta B_{||}^2$, and $\delta B_{xy,\perp}^2$. This is done in Figures \ref{fig:spectra}$c$ and \ref{fig:spectra}$d$ for the cases $m_i/m_e=25$ and 64, respectively. These figures show that $\delta B_z^2$ and $\delta B_{xy,\perp}^2$ are comparable within the power-law tails. Since $\delta B_z^2$ and $\delta B_{xy,\perp}^2$ are expected to be dominated by the OIF and FM/W modes, respectively, this result suggests that these two modes contribute similar amount of energy to the power-law tail. \newline

\noindent A remaining question is whether the presented scenario is plausible in the more realistic case with $m_i/m_e=1836$, where a larger scale separation is expected between $R_{L,i}$ and $R_{L,e}$. We explore this question using Figure \ref{fig:growthrates}$b$, where we plot the growth rate $\gamma$ as a function of $kR_{L,e}$ for OIF and FM/W modes (green and black lines, respectively). This is done assuming $\beta_i=\beta_e=10$, $\Delta p_e=\Delta p_i$, and in two regimes: $i)$ $m_i/m_e=64$ with $\Delta p_j$ such that the maximum growth rate $\gamma_{max}= \omega_{c,e}/7200$ (solid lines), and $ii)$ $m_i/m_e=1836$ with $\Delta p_j$ such that $\gamma_{max}= \omega_{c,e}/10^{6}$ (dotted lines). Since for each value of $\gamma$ there are multiple OIF and FM/W wavevectors ${\bf k}$, in Figure \ref{fig:growthrates}$b$ we chose the maximum and minimum $|{\bf k}|=k$ for each $\gamma$. This way we explore the possibility that the modes with a given $\gamma$ could have wavelengths close to both the electron and ion Larmor radii. We obtained that:
\begin{enumerate}
\item In the case  with $m_i/m_e=64$ and $\gamma_{max}= \omega_{c,e}/7200$, the fastest growing OIF and FM/W modes have $0.03 \lesssim kR_{L,e} \lesssim 0.06$. This range roughly coincides with the peak of the spectra shown in Figures \ref{fig:spectra}$a$ and \ref{fig:spectra}$b$. 
\item In the more realistic case with $m_i/m_e=1836$ and $\gamma_{max}= \omega_{c,e}/10^{6}$, the fastest growing OIF and FM/W modes appear at $0.006 \lesssim kR_{L,e} \lesssim 0.01$. This implies that both modes have a factor $\sim 6$ larger wavelengths than in the $m_i/m_e=64$, $\gamma_{max}= \omega_{c,e}/7200$ case, which is expected because of the factor $\sim 6$ increase in the ratio $R_{L,i}/R_{L,e}$.   
\end{enumerate}   
Thus, for $m_i/m_e=64$ the scale separation between $R_{L,i}$ and $R_{L,e}$ allows the generation of magnetic fluctuations at electron scales with enough energy to pitch-angle scatter the electrons. This result relies on the existence of a power cascade with $d\delta B^2/d\textrm{ln}(k) \propto k^{-2.8}$, which is observed in Figures \ref{fig:spectra}$a$ and \ref{fig:spectra}$b$. However, when $m_i/m_e=1836$ this scenario seems less likely. Indeed, at electron scales ($kR_{L,e} \sim 0.2$) the cascade of OIF modes can produce an amount of energy $\sim (0.2/0.01)^{2.8} \sim 4400$ times smaller than at the ion scales ($kR_{L,i} \sim 0.01$; see dotted-green line in Figure \ref{fig:growthrates}$b$). However, at saturation the OIF and OEF modes are expected to satisfy $\delta B^2/B^2 \propto (s/\omega_{c,j})^{1/2}$ ($j=i$ and $e$, respectively). Thus, in order to provide enough ion and electron pitch-angle scattering, the energy at electron scales should be a factor $\sim (\omega_{c,e}/\omega_{c,i})^{1/2}\sim 1836^{1/2}\sim 43$ smaller than at ion scales, which is $\sim 100$ times larger than what can be produced through the cascade of OIF modes.\footnote{We did not include in this analysis the possible cascade of FM/W waves since, for realistic values of $s/\omega_{c,i}$, their amplitude $\delta B^2/B^2$ ($\propto s/\omega_{c,i}$) should be much smaller than the one of the OIF modes ($\delta B^2/B^2 \propto (s/\omega_{c,j})^{1/2}$). Also, we did not include the possibility of electron scattering via cyclotron resonances (for which $\delta B^2/B^2 \propto s/\omega_{c,e}$), since we do not expect the cascade of OIF or FM/W modes to produce waves with the right polarization to resonate with electrons.}\newline

\noindent This difficulty may get ameliorated if the power cascade process were further modified when $m_i/m_e=1836$, or if in 3D the spectral index of the cascade power-law tail were different from the one obtained in our 2D simulations. Unfortunately, our current simulations can not clarify this aspect of the interplay between the electrons and the OIF and FM/W instabilities. It is important to point out, however, that in realistic settings we do not expect the OEF modes to produce the necessary electron-scale fluctuations either, since our linear calculations show that these modes are stable for the electron anisotropy set by the OIF and FM/W instabilities (with $\Delta p_e=\Delta p_i$). Thus it seems likely that the electron anisotropy should continue to be determined by the OIF and FM/W marginal stability condition with $\Delta p_e=\Delta p_i$.\newline 

\section{Viscous Heating}
\label{sec:viscosity}

\begin{figure}[t!]  
\centering 
\includegraphics[width=8cm]{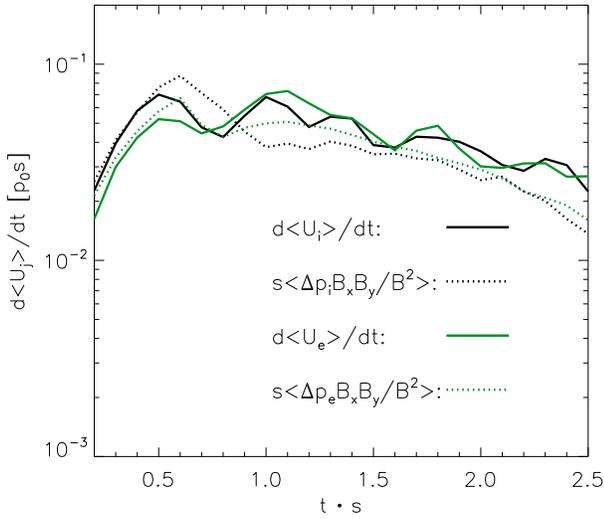} 
\caption{The volume-averaged ion (solid-black) and electron (solid-green) heating rates for run F2. The heating rate by anisotropic viscosity predicted by Equation \ref{eq:anisoheat} for ions and electrons are shown with the dotted-black and the dotted-green lines, respectively. All quantities are normalized by $p_0s$, where $p_0$ is the initial particle pressure in the simulation. For both species there is reasonably good agreement between the particle heating in the simulation and the contribution from the anisotropic stress.}  
\label{fig:energygainfirehose} 
\end{figure}

\noindent The existence of electron and ion pressure anisotropies in general implies the presence of non-diagonal terms in the pressure tensor, which give rise to an effective viscosity for both species. In our case, particle velocities are nearly gyrotropic with respect to $\langle \textbf{\textit{B}} \rangle$, so the relevant pressure tensor component is $p_{xy,j} \propto (p_{\perp,j} - p_{\parallel,j}) B_x B_y/B^2$. It can be shown that this pressure component can tap into the velocity shear of the plasma, producing an increase in the internal energy of the particles. In our case, assuming no heat flux, the internal energy density of species $j$, $U_j$ ($=p_{\perp,j}+p_{\parallel,j}/2$), evolves as \citep{Kulsrud1983,SnyderEtAl1997,SharmaEtAl07}:
\begin{equation}
\frac{\partial U_j}{\partial t} = -s\Delta p_j B_xB_y/B^2 = q\Delta p_j,
\label{eq:anisoheat}
\end{equation}
where $q=-sB_xB_y/B^2$ corresponds to the growth rate of $B$. In the present context, both $q$ and $\Delta p_j$ are negative, which implies an increase in $U_j$. Before the onset of the instabilities, this process is adiabatic and therefore it is a reversible energy gain (in the sense that the increase in $U_j$ would be reverted by reversing the direction of the plasma shear velocity). Indeed, as shown in Figures \ref{fig:cgl_3600and7200}$a$ and \ref{fig:cgl_3600and7200}$b$ for the case of electrons, the early evolution of $p_{\perp,j}$ and $p_{\parallel,j}$ follows the CGL or double adiabatic behavior with $p_{\perp,j} \propto B$ and $p_{\parallel,j} \propto 1/B^2$ (which gives rise to a net growth of $U_j=p_{\perp,j}+p_{\parallel,j}/2$ since the $p_{\parallel,j}$ growth occurs faster than the decrease in $p_{\perp,j}$). Thus, only after the instabilities start keeping $\Delta p_j/p_{\parallel,j}$ in a quasi-stationary regime by breaking $\mu_j$ invariance (after $t\cdot s \approx 0.7$ in Figures \ref{fig:cgl_3600and7200}$a$ and \ref{fig:cgl_3600and7200}$b$), the increase in $U_j$ can be considered as irreversible heating. Also it is important to point out that the role of the instabilities after $t\cdot s \approx 0.7$ is not the direct heating of the particles by wave-particle interactions. Instead, the role of the instabilities is to limit the pressure anisotropy and, therefore, to regulate the viscous heating provided by Equation \ref{eq:anisoheat}.\newline

\noindent Figure \ref{fig:energygainfirehose} quantifies the importance of this heating mechanism by showing the volume-averaged ion (solid-black) and electron (solid-geen) heating rates for run F2. We also show the heating rate by anisotropic viscosity predicted by Equation \ref{eq:anisoheat} for ions (dotted-black) and electrons (dotted-green). For both species there is reasonably good agreement between the particle heating in the simulation and the contribution from the anisotropic stress. This shows that anisotropic viscosity contributes most of the ion and electron heating in collisionless plasmas (with $T_e \sim T_i$), both in the case of decreasing magnetic field presented in this work, as well as in the growing field regime shown in  \cite{RiquelmeEtAl2016}. \newline

\section{Electron Mean Free Path}
\label{sec:conduct}
\noindent Besides regulating the effective plasma viscosity, pitch-angle scattering by velocity-space instabilities is also expected to limit the mean free path of the particles. To quantify this effect we used the distance $D_j(t)$ traveled along $\langle \textbf{\textit{B}}\rangle $ by $2\times 10^4$ ions and electrons \footnote{$D_j(t)\equiv \int_{0}^t \textbf{\textit{v}}_j\cdot \textbf{\textit{B}}/B dt$, where $\textbf{\textit{v}}_j$ is the particle's velocity.}. We calculated their mean free paths assuming that $\langle D_j^2\rangle = tv_{th,j}\langle \lambda_j \rangle$ (where $\langle \lambda_j \rangle$ represents the average mean free path over species $j$, while $v_{th,j}=(k_BT_j/m_j)^{1/2}$ is their thermal speed), which is valid if the particles move diffusively. This allows us to estimate the average mean free path of species $j$ as $\langle \lambda_j \rangle = d\langle D_j^2\rangle/dt/v_{th,j}$.\newline 

\begin{figure}[t!] \centering \includegraphics[width=7.4cm]{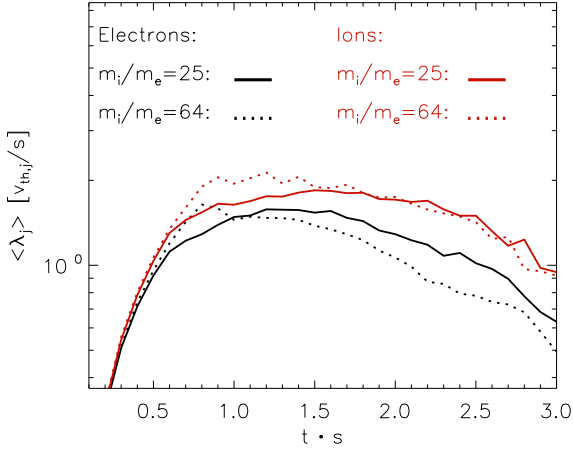} 
\caption{Electron (black) and ion (red) mean free paths (normalized by $v_{th,j}/s$), calculated via the time derivative of the mean squared distance traveled by particles along $\langle \textbf{\textit{B}}\rangle $ ($\langle \lambda_j \rangle = d\langle D_j^2\rangle/dtv_{th,j}$).  Results correspond to runs with $m_i/m_e=64$ (dotted lines; run F1) and $m_i/m_e=25$ (solid lines; run F2).  At early times the particles stream freely, $d\langle D_j^2\rangle/dt \propto t$.   After the velocity-space instabilities saturate, pitch-angle scattering leads to saturation of the mean free path.}  \label{fig:mfpwces7200_mime25and64} \end{figure} 

\noindent Figure \ref{fig:mfpwces7200_mime25and64} shows our estimates of the electron (black) and ion (red) mean free paths for two simulations with $m_i/m_e=25$ and $64$ (runs F2 and F1, respectively), normalized by $v_{th,j}/s$. We see that in both cases there is an initial period when $\langle \lambda_{j} \rangle/(v_{th,j}/s)$ increases as $\sim 2t\cdot s$, which is consistent with the free streaming behavior $\langle D_j \rangle \approx tv_{th,j}$. This is followed by the saturation of  $\langle \lambda_{j} \rangle$, expected to start after a time of the order of the collision time of particles. The behaviors of $\langle \lambda_{e} \rangle$ and $\langle \lambda_{i} \rangle$ in this stage are quite similar, with $\langle \lambda_{e}\rangle/(v_{th,e}/s)$ being somewhat smaller than $\langle \lambda_{i}\rangle/(v_{th,i}/s)$ (by a factor $\sim 1.5$). \newline

\noindent The evolutions of $\langle \lambda_j \rangle$ seen in Figure \ref{fig:mfpwces7200_mime25and64} are expected to be influenced by the pitch-angle scattering of the particles caused by the velocity-space instabilities. In \cite{RiquelmeEtAl2016} we estimated this effect assuming an incompressible fluid with no heat flux, where the scattering produced by the instabilities on species $j$ was incorporated using an effective scattering rate $\nu_{eff,j}$ \citep{Kulsrud1983,SnyderEtAl1997}. In this model, valid for $\Delta p_j/p_{||,j} \ll 1$, $\langle \lambda_j \rangle$ behaves as:
\begin{equation}
\langle \lambda_j \rangle \approx \frac{v_{th,j}}{\nu_{eff, j}} \approx 0.3\frac{v_{th,j}}{q}\frac{\Delta p_j}{p_{||,j}}.
\label{eq:scaling}
\end{equation}

\begin{figure}[t!]  
\centering 
\includegraphics[width=8.7cm]{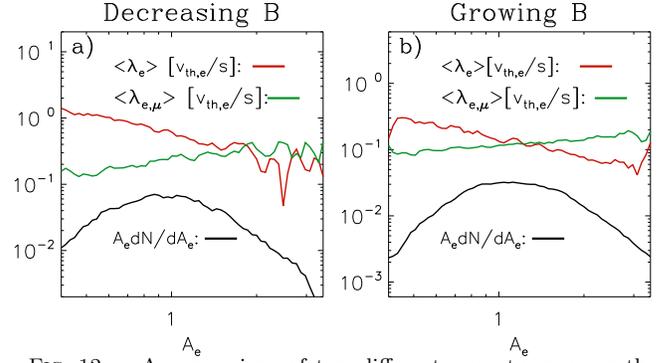} 
\caption{A comparison of two different ways to measure the electron mean free path for the simulations F1 of this paper (panel $a$, with decreasing $B$) and MW1 of \cite{RiquelmeEtAl2016} (panel $b$, with growing $B$), between $t\cdot s=1$ and 3. One way is based on the distance $D_e$ travelled by the electrons along $\textbf{\textit{B}}$, $\langle \lambda_{e} \rangle$ (red), and the other is based on the change of $\mu_e$, $\langle \lambda_{e,\mu} \rangle$ (green). (See footnote \ref{foot} for details on the calculation of $\langle \lambda_{e} \rangle$ and $\langle \lambda_{e,\mu} \rangle$). The mean free paths are plotted as a function of $A_e\equiv \langle v_{\perp,e}^2\rangle/\langle v_{\parallel,e}^2\rangle$, where $\langle \rangle$ represents the time average between $t\cdot s=1$ and 3 for each electron. In panels $a$ and $b$, the black lines show the respective distributions of $A_e$, $A_edN/dA_e$.}  
\label{fig:mfpmu_paper} 
\end{figure}

\noindent Equation \ref{eq:scaling} provides a good approximation to the mean free path of particles in the case of growing magnetic fields \citep{RiquelmeEtAl2016}, where $\langle \lambda_j \rangle$ is regulated by the whistler and mirror instabilities. In the cases of decreasing magnetic field, however, this is not the case. Using our measurements of $\Delta p_j/p_{||,j}$ for runs F2 and F1 (see Figures \ref{fig:benandanis_wces2500_mime25and64}$d$ and \ref{fig:benandanis_wces2500_mime25and64}$e$) one obtains that, at $t\cdot s=1.5$, $\langle \lambda_j \rangle \approx 0.15 v_{th,j}/s$. However, Figure \ref{fig:mfpwces7200_mime25and64} shows that the average mean free paths of ions and electrons are $\sim 10$ times larger than this simple estimate. On the other hand, considering the evolution of $\Delta p_j/p_{||,j}$ and $B^2/B_xB_y$ (needed to determine $q$) from $t\cdot s =1.5$ to $t\cdot s =3$, $\langle \lambda_i \rangle$ should decrease by a factor $\sim 2$, which is essentially what Figure \ref{fig:mfpwces7200_mime25and64} shows. Thus, putting aside the factor $\sim 10$ difference, the scaling of $\langle \lambda_j \rangle$ on $\Delta p_j/p_{||,j}$ and $q$ presented in Equation \ref{eq:scaling} is reasonably well reproduced by the simulations with decreasing $B$.\newline

\noindent The factor $\sim 10$ discrepancy is interesting, partly because the behavior of $\nu_{eff,j}$ suggested by Equation \ref{eq:scaling} ($\nu_{eff,j}\approx 3qp_{||,j}/\Delta p_j$) is well reproduced in the case of ions in previous hybrid-PIC simulations that studied the saturated state of the firehose and mirror instabilities \citep{Kunz2014}. In that case, however, $\nu_{eff,i}$ is not measured using $D_i$. Instead, they constructed a distribution of the times taken by each ion to change its $\mu_i$ by a factor of {\it e}, and then approximated $\nu_{eff,i}^{-1}$ by the width of the distribution. Thus, in order to clarify this discrepancy, we compared two different measurements of the electron mean free path from run F1, one using the variations of $\mu_e$ (we will refer to this estimate as $\langle \lambda_{e,\mu} \rangle$) and the other one using $D_e$ ($\langle \lambda_{e} \rangle$). These measurements of $\langle \lambda_e \rangle$ and $\langle \lambda_{e,\mu} \rangle$ are not defined for different times (as in Figure \ref{fig:mfpwces7200_mime25and64}), but they correspond to averages between $t\cdot s=1$ and 3.\footnote{For a given electron population, $\langle \lambda_{e} \rangle$ is estimated by first calculating for each electron the quantity $\lambda_{e}=(D(t\cdot s=3)^2-D(t\cdot s=1)^2)/(\langle|v_{||,e}|\rangle\Delta t)$ (where $\Delta t$ is simply the time elapsed between $t\cdot s=1$ and 3 and $\langle |v_{||,e}| \rangle$ is the average magnitude of the electron velocity parallel to $\textbf{\textit{B}}$ in the same period), and then taking the average over the population. For $\langle \lambda_{e,\mu}\rangle$, we construct a distribution of the times taken by each electron to change $\mu_e$ by a factor {\it e} with respect to its value at $t\cdot s=1$, and then $\langle \lambda_{e,\mu} \rangle$ is estimated multiplying the width of the distribution by $v_{th,e}$.\label{foot}} The comparison was made for different groups of electrons, defined by the parameter $A_e\equiv \langle v_{\perp,e}^2\rangle/\langle v_{\parallel,e}^2\rangle$, where $\langle \rangle$ represents the average between $t\cdot s=1$ and 3 for each electron. We use $A_e$ as a way to quantify the pitch-angle of electrons, which, as we will see below, affects the behaviors of $\langle \lambda_{e} \rangle$ and $\langle \lambda_{e,\mu} \rangle$ in different ways. \newline

\noindent Our results are shown in Figure \ref{fig:mfpmu_paper}$a$. We see that $\langle \lambda_{e} \rangle$ and $\langle \lambda_{e,\mu} \rangle$ roughly coincide for $A_e \gtrsim 1$ (pitch-angle $\gtrsim 45^o$). However, for $A_e \lesssim 1$ (pitch-angle $\lesssim 45^o$), $\langle \lambda_{e} \rangle$ becomes significantly larger than $\langle \lambda_{e,\mu} \rangle$. This is consistent with the fact that, for electrons with small pitch-angle, an order unity variation in $v_{\perp,e}^2$ due to scattering (which implies an order unity variation in $\mu_e$) does not imply that they reverse their velocity along $\textbf{\textit{B}}$. Also, when averaged over the entire $A_e$ distribution (shown by the black line in Figure \ref{fig:mfpmu_paper}$a$), $\langle \lambda_{e} \rangle$ is a factor $\sim 10$ larger than $\langle \lambda_{e,\mu} \rangle$, showing that using $\langle \lambda_{e,\mu} \rangle$ would essentially eliminate the discrepancy between our estimated mean free path and Equation \ref{eq:scaling}.\footnote{The factor $\sim 10$ diference between our measured $\langle \lambda_{e} \rangle$ and the estimate given by Equation \ref{eq:scaling} was also obtained in simulations with infinite mass ions, where the electron scattering is dominated by the OEF modes. Also, the same difference is obtained in simulations with infinite mass ions and initial $\beta_e=5$, suggesting that this discrepancy is fairly insensitive to $\beta_e$, at least in the moderate $\beta_e$ regime that we studied.}\newline

\noindent Given this difference between $\langle \lambda_{e,\mu} \rangle$ and $\langle \lambda_{e} \rangle$, it is important to understand why in the case of growing magnetic field studied by \cite{RiquelmeEtAl2016} the behavior of $\langle \lambda_{e} \rangle$ is well reproduced by Equation \ref{eq:scaling}. Figure \ref{fig:mfpmu_paper}$b$ shows the same quantities as Figure \ref{fig:mfpmu_paper}$a$ but for run MW1 of \cite{RiquelmeEtAl2016}. We see that the trend for $\langle \lambda_{e} \rangle$ to grow relative to $\langle \lambda_{e,\mu} \rangle$ as $A_e$ decreases is maintained in this case. However, for all values of $A_e$, the growing $B$ case tends to have a smaller ratio $\langle \lambda_e \rangle / \langle \lambda_{e,\mu} \rangle$ compared to the decreasing $B$ case, making  $\langle \lambda_e \rangle \sim \langle \lambda_{e,\mu} \rangle$ if the average over the whole distribution of $A_e$ (black line) is considered. \newline
\begin{figure}[t!]  
\centering 
\includegraphics[width=9.5cm]{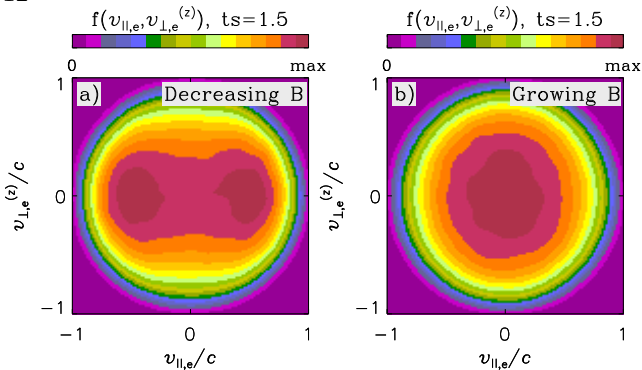} 
\caption{Panels $a$ and $b$ show the electron velocity distribution $f(v_{||,e},v_{\perp,e}^{(z)})$ at $t\cdot s=1.5$ for run F1 (decreasing $B$) and MW1 \citep[growing $B$;][]{RiquelmeEtAl2016}, respectively ($v_{||,e}$ and $v_{\perp,e}^{(z)}$ are respectively the electron velocity parallel to $\langle \textbf{\textit{B}}\rangle$ and to the $z$ axis, which are mutually perpendicular). Run F1 shows that, for $v_e \lesssim 0.6c$ ($v_e \equiv (v_{||,e}^2+v_{\perp,e}^{(z)2})^{1/2}$), the distribution is dominated by electrons with small pitch-angle, while in run MW1 the velocity distribution appears more similar to a bi-Maxwellian distribution with $p_{\perp,e}>p_{||,e}$.}  
\label{fig:veldist} 
\end{figure}

\noindent This difference between the growing and decreasing $B$ cases appears to be due to the specific effect of the relevant instabilities on the electron velocities. This is suggested by Figure \ref{fig:veldist}$a$, which shows the electron velocity distribution $f(v_{||,e},v_{\perp,e}^{(z)})$ for run F1 at $t\cdot s=1.5$ (corresponding to the saturated stage of the FM/W and OIF instabilities), where $v_{||,e}$ and $v_{\perp,e}^{(z)}$ are respectively the electron velocity parallel to $\langle \textbf{\textit{B}} \rangle$ and to the $z$ axis (which are mutually perpendicular). We see that, for $v_e \lesssim 0.6c$ ($v_e \equiv (v_{||,e}^2+v_{\perp,e}^{(z)2})^{1/2}$), $f(v_{||,e},v_{\perp,e}^{(z)})$ is dominated by electrons with rather small pitch-angle ($\lesssim 45^o$). This suggests that for $v_e \lesssim 0.6c$, the scattering process occurs in a way that disfavors the diffusion of electrons towards smaller values of $|v_{||,e}|$, which in turn precludes the reversal of $v_{||,e}$, contributing to increasing $\langle \lambda_e \rangle$. For comparison, in Figure \ref{fig:veldist}$b$ we show the analogous electron distribution for run MW1 of \cite{RiquelmeEtAl2016} (growing $B$) at $t\cdot s=1.5$, where $f(v_{||,e},v_{\perp,e}^{(z)})$ appears more similar to a bi-Maxwellian distribution with $p_{\perp,e} > p_{||,e}$. Notice that the dominance of small pitch-angle electrons for $v_e \lesssim 0.6c$ seen in the decreasing field case is similar to the modification to the ion velocity distribution found by previous hybrid-PIC simulations of an expanding box, where the ion scattering is also provided by the FM/W and OIF instabilities \citep{MatteiniEtAl2006, HellingerEtAl2008}.\newline

\noindent In summary, both for growing and decreasing $B$, the estimate of the electron mean free path provided by Equation \ref{eq:scaling} is fairly well reproduced by $\langle \lambda_{e,\mu} \rangle$ (as it was shown by \cite{Kunz2014} in the case of ions). However, since $\langle \lambda_{e} \rangle$ is based on the direct calculation of the distance travelled by the electrons along $\textbf{\textit{B}}$, this quantity provides a more meaningful measurement of the electron mean free path for the purpose of quantifying the thermal conductivity of the plasma. In the decreasing field case, $\langle \lambda_{e} \rangle$ is a factor $\sim 10$ larger than $\langle \lambda_{e,\mu} \rangle$, most likely due to the specific electron scattering mechanism provided by the FW/W and OIF instabilities. Thus, in this case the electron mean free path is best described by the relation:
\begin{equation}
\langle \lambda_j \rangle \approx 3\frac{v_{th,j}}{q}\frac{\Delta p_j}{p_{||,j}},
\label{eq:scaling2}
\end{equation} 
which is valid for $2\lesssim \beta_e \lesssim 20$, and only differs from Equation \ref{eq:scaling} by its prefactor $\sim 3$ instead of $\sim 0.3$.\newline

\section{Discussion and Implications}
\label{sec:conclu}

\noindent We used particle-in-cell (PIC) plasma simulations to study the nonlinear, saturated stage of various ion and electron velocity-space instabilities relevant for collisionless plasmas. We focused on instabilities driven by pressure anisotropy with $p_{\perp,j} < p_{||,j}$. To capture the nonlinear regime in a self-consistent way, we imposed a shear velocity in the plasma, which decreases the background magnetic field. This drives $p_{\perp,j} < p_{||,j}$ due to the adiabatic invariance of the magnetic moment (the driving timescale is much longer than the gyroperiod of the particles). This, in turn, drives velocity-space instabilities, which inhibit the growth of pressure anisotropy. The relevant instabilities in this regime, as suggested by linear theory, are: $i)$ the purely growing oblique ion-firehose (OIF) and the resonant fast magnetosonic/whistler (FM/W) modes, which are mainly driven by the ions, and $ii)$ the purely growing oblique electron-firehose (OEF) and the resonant Alfv\'en/ion-cyclotron (A/IC) modes, which are driven by the electrons. The nonlinear state of these instabilities is expected to be influenced by the simultaneous presence of ion and electron anisotropies on the different modes. In order to achieve reasonable scale separation between these modes, we mainly used $m_i/m_e=25$ and 64. Our results, valid for the regime $2\lesssim \beta_i \approx \beta_e \lesssim 20$, showed no significant difference between these two mass ratios.\newline

\noindent We found that the mechanism for regulating the ion and electron anisotropies consists in the growth of OIF and FM/W modes, which affect equally the ions and electrons, rendering $\Delta p_e \approx \Delta p_i$. The numerically obtained ion and electron anisotropies are well approximated by the linear threshold for the growth of the OIF and FM/W modes with $\Delta p_e = \Delta p_i$ and with growth rate $\sim s$. The electron pressure anisotropy in simulations with infinite mass ions (where the ions only provide a neutralizing charge) is dominated by the OEF and A/IC modes, giving a factor $\sim 2$ larger anisotropy than in the cases with finite $m_i/m_e$. We attribute this result to the rather strong destabilizing effect of the electron pressure anisotropy, $\Delta p_e$, on the OIF and FM/W modes (as already suggested by previous linear dispersion analyses \cite{MichnoEtAl2014, ManevaEtAl2016}), which in turn maintains $\Delta p_e$ at a value significantly lower than the one necessary to make the OEF and A/IC modes grow at a rate $\sim s$.\newline

\noindent Although the amplitude of the OIF and FM/W modes depend on the ratio $\omega_{c,i}/s$, the value of the parameters $m_i/m_e$ and $\omega_{c,e}/s$ used in the simulations do not affect our conclusions. Also, based on our linear Vlasov calculations \citep{Verscharen2013}, we infer that the presented scenario should hold in the $m_i/m_e=1836$, highly magnetized ($\omega_{c,i}/s \gg 1$) case relevant for real astrophysical plasmas. However, an important point that our simulations could not completely clarify (due to the lack of sufficient ion and electron scale separation) is the mechanism by which the electrons would be pitch-angle scattered in the saturated stage of the OIF and FM/W instabilities in the case of $m_i/m_e=1836$. Answering this question requires using significantly larger mass ratios and magnetizations (and possibly 3D runs), so we differ this aspect of the study to a future work.\newline

\noindent We have also used our simulations to verify the expected viscous heating of particles, which is described in Equation \ref{eq:anisoheat}, and arises due to pressure anisotropies tapping into the free energy in the shear motion of the plasma. Figure \ref{fig:energygainfirehose} shows a good agreement between the heating of the particles in our simulations and the expectation from Equation \ref{eq:anisoheat}. This result is valid for decreasing magnetic fields, as shown here, and for growing fields as shown by \cite{RiquelmeEtAl2016}.\newline

\noindent With the intention to quantify the thermal conductivity in these plasmas, we have also computed the mean free path of species $j$, $\langle \lambda_j \rangle$, during the nonlinear stage of the OIF and FM/W instabilities. The average mean free path of both ions and electrons is reasonably well described by Equation \ref{eq:scaling2}. The scaling factors in this equation are the same as in Equation \ref{eq:scaling}, which is based on a model where the mean free path of species $j$ is determined by an effective scattering rate $\nu_{eff,j}$ that sets the rate at which the $\mu_j$ invariance is broken \citep{Kulsrud1983,SnyderEtAl1997}. However, the prefactor in Equation \ref{eq:scaling2} is $\sim 10$ times larger than in the case of Equation \ref{eq:scaling}. 

We explained this discrepancy for the case of the electrons by noticing that the variation of $\mu_e$ provides a good estimate of $\langle \lambda_e \rangle$ only for electrons with relatively large pitch-angle ($\gtrsim 45^o$). For electrons with pitch-angles smaller than $\sim 45^o$, $\langle \lambda_e \rangle$ can become a factor $\sim 10$ larger. This is in contrast with the case of growing $B$ \citep{RiquelmeEtAl2016}, where Equation \ref{eq:scaling} provides a reasonably good estimate for $\langle \lambda_j \rangle$, despite the fact that $\langle \lambda_j \rangle$ also grows as the pitch-angle decreases. This difference is likely due to the specific electron scattering mechanism provided by the FW/W and OIF instabilities, which tends to preclude the reversal of $v_{||,e}$, contributing to increase $\langle \lambda_e \rangle$.\newline 

\noindent The results shown in this work, as well as those presented in \cite{RiquelmeEtAl2016} for the growing $B$ case, are relevant for quantifying the viscous heating and thermal conductivity in various low-collisionality astrophysical plasmas, including low-luminosity accretion flows around compact objects, the ICM, and the heliosphere.\newline 

\acknowledgements
This work was supported by NSF grants AST 13-33612 and 1715054, a Simons Investigator Award to EQ from the Simons Foundation and the David and Lucile Packard Foundation. DV also acknowledges support from NSF/SHINE grant AGS-1460190, NSF/SHINE grant AGS-1622498, NASA grant NNX16AG81G, and a UK STFC Ernest Rutherford Fellowship (ST/P003826/1). We are also grateful to the UC Berkeley-Chile Fund for support for collaborative trips that enabled this work. This work used the Extreme Science and Engineering Discovery Environment (XSEDE), which is supported by National Science Foundation grant number ACI-1053575.



\begin{thebibliography}{99}
\addcontentsline{toc}{section}{Bibliography}
\bibitem[Buneman(1993)]{Buneman93} Buneman, O. 1993, ``Computer Space Plasma Physics'', Terra Scientific, Tokyo, 67
\bibitem[Camporeale et al.(2008)]{CamporealeEtAl2008} Camporeale, E. \& Burgess, D. 2008, J. Geophys. Res., 113, A07107
\bibitem[Camporeale et al.(2010)]{CamporealeEtAl2010} Camporeale, E. \& Burgess, D. 2010, ApJ, 710, 1848
\bibitem[Chew et al.(1956)]{ChewEtAl56} Chew, G.F., Goldberger, M.L., \& Low, F.E. 1956, Proc. R. Soc. A 236, 112
\bibitem[Gary et al.(1998)]{GaryEtAl1998} Gary, S. P., Li, H., O'Rourke, S. \& Winske, D. 1998, J. Geophys. Res., 103, 14,567
\bibitem[Kulsrud(1983)]{Kulsrud1983} Kulsrud, R. M. 1983, in Handbook of Plasma Physics, ed. M. N. Rosenbluth \& R. Z. Sagdeev (Amsterdam: North Holland), 115
\bibitem[Hellinger et al.(2000)]{HellingerEtAl2000} Hellinger, P. \& Matsumoto, H. 2000, J. Geophys. Res., 105, 10,519
\bibitem[Hellinger et al.(2008)]{HellingerEtAl2008} Hellinger, P. \& Travnicek, P. 2008,  J. Geophys. Res., 113, A10109
\bibitem[Hellinger et al.(2014)]{HellingerEtAl2014} Hellinger, P., Travnicek, P., Decyk, V., \& Schriver, D. 2014,  J. Geophys. Res., 119, 59
\bibitem[Komarov et al.(2016)]{KomarovEtAl16} Komarov, S. V., Churazov, E. M., Kunz, M. W., \& Schekochihin, A. A. 2016, MNRAS, 460, 467
\bibitem[Kunz et al.(2014)]{Kunz2014} Kunz, M. W., Schekochihin, A., \& Stone, J.~M.\ 2014, Physical Review Letters, 112, 205003 
\bibitem[Kunz et al.(2015)]{Kunz2015} Kunz, M. W., Schekochihin, A., Chen, C.,  Abel, I., \& Cowley, S. 2017, JPP, 81, 325810501 
\bibitem[Li et al.(2000)]{LiEtAl2000} Li, X. \& Habbal, S. R. 2000, J. Geophys. Res., 105, 27377
\bibitem[Lyutikov(2007)]{Lyutikov07} Lyutikov, M. 2007, \apj, 668, L1
\bibitem[Maneva et al.(2016)]{ManevaEtAl2016} Maneva, Y., Lazar, M., Vi\~{n}as, A., \& Poedts, S. 2016, \apj, 832, 64
\bibitem[Marsch (2006)]{Marsch2006} Marsch, E. 2006, Living Rev. Solar Phys., 3, 1
\bibitem[Maruca et al.(2011)]{MarucaEtAl11} Maruca, B. A., Kasper, J. C. \& Bale, S. D. 2011, Phys. Rev. Lett. 107, 201101
\bibitem[Matteini et al.(2006)]{MatteiniEtAl2006} Matteini, L., Landi, S., Hellinger, P., \& Velli, M. 2006,  J. Geophys. Res., 111, A10101
\bibitem[Michno et al.(2014)]{MichnoEtAl2014} Michno, M., Lazar, M., Yoon, P., \& Schlickeiser, R. 2014, \apj, 781, 49
\bibitem[Quest et al.(1996)]{QuestEtAl1996} Quest, K. B. \& Shapiro, V. D. 1996, J. Geophys. Res., 101,24,457
\bibitem[Remya et al.(2013)]{RemyaEtAl13} Remya, B., Reddy, R. V., Tsurutani, B. T., Lakhina, G. S., \& Echer, E. 2013, JGRA, 118, 785
\bibitem[Riquelme et al.(2015)]{RiquelmeEtAl2015} Riquelme, M. A., Quataert, \& Verscharen, D. 2015, \apj, 800, 27
\bibitem[Riquelme et al.(2016)]{RiquelmeEtAl2016} Riquelme, M. A., Quataert, \& Verscharen, D. 2016, \apj, 824, 123
\bibitem[Schekochihin et al.(2005)]{SchekochihinEtAl05} Schekochihin, A. A., Cowley, S. C., Kulsrud, R. M., Hammett, G. W., \& Sharma, P. 2005, \apj, 629, 139
\bibitem[Sharma et al.(2006)]{SharmaEtAl06} Sharma, P., Hammett, G. W., Quataert, E., \& Stone, J. 2006, \apj, 637, 952
\bibitem[Sharma et al.(2007)]{SharmaEtAl07} Sharma, P., Quataert, E., Hammett, G. W., \& Stone, J. 2007, \apj, 667, 714
\bibitem[Snyder et al.(1997)]{SnyderEtAl1997} Snyder, P. B., Hammett, G. W., \& Dorland, W. 1997, Phys. Plasmas, 4, 3974
\bibitem[Spitkovsky(2005)]{Spitkovsky05} Spitkovsky, A. 2005, AIP Conf. Proc, 801, 345, astro-ph/0603211
\bibitem[Squire et al.(2017)]{SquireEtAl2017} Squire, J., Kunz, M., Quataert, E., \& Schekochihin, A. 2017,  arXiv:1705.01956
\bibitem[Verscharen et al.(2013)]{Verscharen2013} Verscharen, D., Bourouaine, S., Chandran, B.~D.~G., \& Maruca, B.~A. 2013, \apj, 773, 8 
\bibitem[Verscharen et al.(2017)]{Verscharen2017} Verscharen, D., Chen, C., \& Wicks, R. 2017, \apj, 840, 106
\end{thebibliography}
\end{document}